\documentclass{aa}   % two-column printer version

\usepackage{graphicx}
\usepackage{txfonts}
\usepackage[breaklinks=true,colorlinks=true,linkcolor=blue,citecolor=gray,urlcolor=magenta]{hyperref}
\usepackage[]{listings} % used to color code
\usepackage{color}
\definecolor{light-gray}{gray}{0.95}
\newcommand{\simgt}{\,\rlap{\lower 3.5 pt \hbox{$\mathchar \sim$}} \raise
1pt \hbox {$>$}\,}
\newcommand{\simlt}{\,\rlap{\lower 3.5 pt \hbox{$\mathchar \sim$}} \raise
1pt \hbox {$<$}\,}

\usepackage{xcolor}
\usepackage{natbib,twoopt}
\bibpunct{(}{)}{;}{a}{}{,}             %% natbib format for A&A and ApJ
\makeatletter
  \newcommandtwoopt{\citeads}[3][][]{\href{http://adsabs.harvard.edu/abs/#3}%
    {\def\hyper@linkstart##1##2{}%
     \let\hyper@linkend\@empty\citealp[#1][#2]{#3}}}
  \newcommandtwoopt{\citepads}[3][][]{\href{http://adsabs.harvard.edu/abs/#3}%
    {\def\hyper@linkstart##1##2{}%
     \let\hyper@linkend\@empty\citep[#1][#2]{#3}}}
  \newcommandtwoopt{\citetads}[3][][]{\href{http://adsabs.harvard.edu/abs/#3}%
    {\def\hyper@linkstart##1##2{}%
     \let\hyper@linkend\@empty\citet[#1][#2]{#3}}}
  \newcommandtwoopt{\citeyearads}[3][][]%
    {\href{http://adsabs.harvard.edu/abs/#3}
    {\def\hyper@linkstart##1##2{}%
     \let\hyper@linkend\@empty\citeyear[#1][#2]{#3}}}
\makeatother
\begin{document}

\title{Three Dimensional Optimal Spectral Extraction (TDOSE) from Integral Field Spectroscopy }
\titlerunning{TDOSE}
\authorrunning{Schmidt et al. (2019)}

%\subtitle{I. Overviewing the $\kappa$-mechanism}

\author{K. B. Schmidt\inst{1}\thanks{E-mail: kbschmidt@aip.de},
L.~Wisotzki\inst{1}, 
T. Urrutia\inst{1}, 
J. Kerutt\inst{1}, 
D. Krajnovi{\'c}\inst{1},
E. C. Herenz\inst{2},
R. Saust\inst{1}, 
% Albhabetic
T. Contini\inst{3},
B. Epinat\inst{3,4},
H. Inami\inst{5,6},
and 
M. V. Maseda\inst{7}
%\NB{et al. TBD}
}

\institute{
$^1$ Leibniz-Institut f\"{u}r Astrophysik Potsdam (AIP), An der Sternwarte 16, 14482, Potsdam, Germany\\
$^2$ Department of Astronomy, Stockholm University, AlbaNova University Centre, 106 91, Stockholm, Sweden\\
$^3$ IRAP, Universit\'e de Toulouse, CNRS, CNES, UPS, (Toulouse), France\\
$^4$ Aix Marseille Univ., CNRS, CNES, LAM, Marseille, France\\
$^5$ Univ. Lyon 1, ENS de Lyon, CNRS, Centre de Recherche Astrophysique de Lyon (CRAL) UMR5574, 69230 Saint-Genis-Laval, France \\
$^6$ Hiroshima Astrophysical Science Center, Hiroshima University, 1-3-1 Kagamiyama, Higashi-Hiroshima, Hiroshima, 739-8526\\
$^7$ Leiden Observatory, Leiden University, P.O. Box 9513, 2300 RA, Leiden, The Netherlands
}
\date{Received 9 May 2019 / Accepted 13 June 2019}

% Abstract of the paper
 \abstract{
% ---------------- Aims ----------------
The amount of integral field spectrograph (IFS) data has grown considerable over the last few decades.
The demand for tools to analyze such data is therefore bigger now than ever. 
We present TDOSE; a flexible Python tool for Three Dimensional Optimal Spectral Extraction from IFS data cubes.
% ---------------- Method ----------------
TDOSE works on any three-dimensional data cube and bases the spectral extractions on morphological reference image models.
By default, these models are generated and composed of multiple multivariate Gaussian components, but can also be constructed with independent modeling tools and be provided as input to TDOSE.
In each wavelength layer of the IFS data cube, TDOSE \emph{simultaneously} optimizes all sources in the morphological model to minimize the difference between the scaled model components and the IFS data.
The flux optimization produces individual data cubes containing the scaled three-dimensional source models.
This allows for efficient de-blending of flux in both the spatial and spectral dimensions of the IFS data cubes, and extraction of the corresponding one-dimensional spectra.
%
% ---------------- Results ----------------
TDOSE implicitly requires an assumption about the two-dimensional light distribution. 
We describe how the flexibility of TDOSE can be used to mitigate and correct for deviations from the input distribution.
Furthermore, we present an example of how the three-dimensional source models generated by TDOSE can be used to improve two-dimensional maps of physical parameters like velocity, metallicity or SFR, when flux contamination is a problem. 
By extracting TDOSE spectra of $\sim$150 [OII] emitters from the MUSE-Wide survey we show that the median increase in line flux is $\sim$5\% when using multi-component models as opposed to single-component models. 
However, the increase in recovered line emission in individual cases can be as much as 50\%.
Comparing the TDOSE model-based extractions of the MUSE-Wide [OII] emitters with aperture spectra, the TDOSE spectra provides a median flux (S/N) increase of 9\% (14\%). 
Hence, TDOSE spectra optimizes the S/N while still being able to recover the total emitted flux.
TDOSE version 3.0 presented in this paper is available at \url{https://github.com/kasperschmidt/TDOSE} and \cite{Schmidt:2019cq}.
}
 

   \maketitle
%
%%%%%%%%%%%%%%%%% BODY OF PAPER %%%%%%%%%%%%%%%%%%
% ======================================================================
\section{Introduction}
\label{sec:intro}

%\CH{This is text that has been changed...}

With the advent of large-area three-dimensional (3D) integral field spectrographs (IFSs) and multi-object spectrographs over the last few decades, larger and larger spectroscopic samples of galaxies have become available. 
In particular with the growth of the field-of-view (FoV) of IFSs, areas of more than 100 square arcminutes now have complete spectroscopic coverage down to the limiting depth of the observations, which are often well below fluxes of 10$^{-17}$erg/s/cm$^2$/\AA.

In particular, the optical Multi Unit Spectroscopic Explorer \citep[MUSE;][]{2014Msngr.157...13B,2010SPIE.7735E..08B} on ESO's Very Large Telescope (VLT) and the Hobby-Eberly Telescope Dark Energy Experiment \citep[HETDEX;][]{2016ASPC..507..393H,2012SPIE.8446E..0NH} are current IFS facilities mapping large areas on the sky more efficiently than has previously been possible.
MUSE, has since its start of operations in 2014 been instrumental in providing sensitive wide-area \citep[][]{2017A&A...606A..12H,2019A&A...624A.141U} and deep pencil-beam surveys \citep{2017A&A...608A...1B,2015A&A...575A..75B} with complete medium-resolution spectroscopic coverage.
Most of these data have been taken over already well-known legacy fields with extensive ancillary photometric data available, but has nevertheless revealed new understanding and insights about the general galaxy population, due to its blind spectroscopic nature.
Among these results, it is worth noting the spectroscopic identification of Ly$\alpha$ emitting galaxies un-detected in even the deepest existing \emph{Hubble Space Telescope} (HST) photometry \citep[][]{2017A&A...608A...2I,2018ApJ...865L...1M}, the discovery of ubiquitous extended Ly$\alpha$ halos in Ly$\alpha$ emitters \citep[][Saust et al., in prep.]{2017A&A...608A...8L,2016A&A...587A..98W,2018Natur.562..229W}, and a likely bias in previous estimates of the faint end of the Ly$\alpha$ luminosity function \citep[LF;][]{2017A&A...608A...6D,2017MNRAS.471..267D,2019A&A...621A.107H}.
These studies, focused on Ly$\alpha$ emission, were all enabled by the wide-area blind IFS searches with MUSE.
But also detailed studies of samples of more nearby galaxies, have become considerably more sophisticated in recent years, thanks to the advancement of the IFS capabilities. 
In particular, the significant progress has been driven by dedicated IFS surveys on individual objects like
SAURON \citep{2002MNRAS.329..513D}, SINS \citep{ForsterSchreiber:2009hm}, ATLAS3D \citep{2011MNRAS.413..813C}, CALIFA \citep{2016A&A...594A..36S,2015A&A...576A.135G,2013A&A...549A..87H,2012A&A...538A...8S}, SAMI \citep{2018MNRAS.481.2299S,2018MNRAS.475..716G,2015MNRAS.446.1567A}, MaNGA \citep{2015ApJ...798....7B}, and KMOS-3D \citep[e.g.][]{2015ApJ...799..209W}.
Similarly, the capability of efficiently surveying large areas on the sky, has become possible by for instance taking advantage of the 3D capabilities of the HST grisms \citep[e.g.;][]{2013MNRAS.432..285S,2016ApJ...828...27N,2016ApJ...817L...9N,2013ApJ...763L..16N,2017ApJ...837..126V,2016ApJ...833..178V,2015ApJ...814..161V,2017ApJ...837...89W}, or by exploiting the IFSs surveying the sky for HETDEX or as part of the numerous MUSE programs currently being carried out.
Taking advantage of the large FoV and increased sensitivity studies with MUSE have presented the metallicity, kinematics, emission line diagnostics, cluster masses, etc. of samples of both nearby and distant objects
\citep[e.g.;][]{2016A&A...591A..49C,
2017A&A...608A...5G,
2017A&A...608A...6D,
2017MNRAS.469.3946L,
2017ApJ...844...48P,
2017MNRAS.467.3140S,
2017A&A...605A.118F,
2018MNRAS.478.4293C,
2018MNRAS.477.5327K,
2018MNRAS.477...18P,
2018A&A...618A..40P,
2018A&A...617A..62F,
2018MNRAS.473..663M}. 
Such studies have only become possible with large complete spectroscopic samples from IFSs.  
And with new IFS capabilities being planned and becoming available on upcoming telescopes including the James Webb Space Telescope (JWST), ESO's Extremely Large Telescope (ELT) and the Thirty Meter Telescope (TMT), the amount of IFS data available shows no signs of stagnation.

As showcased by the studies mentioned above, IFS data are particularly useful for pixel-by-pixel spectral analysis to obtain maps of for instance metallicity, kinematics and ionization parameters. 
But also samples of integrated one-dimensional (1D) spectra of complete flux limited galaxy samples for population statistics like LF analysis, emission line characteristics and comparisons with parameters derived from photometric and spectral galaxy models, are key areas where the large amount of IFS data has been transforming.

Irrespective of whether the science case is focused on resolved maps or integrated 1D spectra, a crucial part of any IFS data extraction, is accounting for contaminating light. Here contamination refers to any light coming from fore- or background objects not part of the object(s) of interest.
To perform such de-blending, i.e. accounting for the flux contribution of all objects to all pixels in the FoV, a spectral extraction accounting for the morphology of individual objects and the wavelength dependent point spread function (PSF) optimizing the signal-to-noise ratio (S/N) of the spectra, is needed. 
Tools for extracting point-sources, e.g., stars \citep[PampelMuse;][]{2013A&A...549A..71K} and extended objects, e.g.,  \citep[AUTOSPEC;][]{2018ApJ...869...68G} have been developed with this in mind. 
Such spectral extractions where the source morphology is accounted for, are often referred to as "optimal" spectral extractions.

In this paper, we present a complementary tool for Three Dimensional Optimal Spectral Extraction (TDOSE) to accommodate the large amount of deep spectroscopic IFS data of galaxies available today.
TDOSE builds on the same principles as the spectral extraction tool for crowded stellar fields, PampelMuse, applied to extensive MUSE data by \cite{2016A&A...588A.148H} and \cite{2016A&A...588A.149K,2018MNRAS.473.5591K}.
However, TDOSE expands these concepts for applicability to non-point sources, i.e. galaxies and other non-stellar objects.
TDOSE performs simultaneous extraction taking both the wavelength dependent PSF and the morphology of individual galaxies in the FoV into account to facilitate de-blending of flux and hence the spectra of neighboring sources.
As we will show neighboring sources do not need to be distinguishable in the IFS data, as long as they are marginally resolved in the reference imaging.   
This represents one of the main advantages of including prior information from ancillary data in the extraction as opposed to only relying on the IFS data alone.

Throughout this paper a \emph{source} refers to a `light source', which does not necessarily correspond to a single object. An \emph{object} corresponds to a collection of sources, such that a multi-component galaxy can be extracted combining the fluxes of multiple sources, e.g., different [OII] regions, spiral features, or extended halos.

The paper is structured as follows: 
In Section~\ref{sec:optimalext} we discuss the term "optimal extractions".
In Section~\ref{sec:tdose} we describe the framework of TDOSE and the individual stages of spectral extractions performed with the software.
We describe the capabilities and main limitation of the software, by examples of spectral extractions from MUSE data cubes in Section~\ref{sec:MUSEextractions}.
This include examples of recovering spectra only partially covered in the IFS data, de-blending sources, comparing extractions based on single-source and multi-source object models \citep[e.g. generated with GALFIT;][]{2010AJ....139.2097P,2002AJ....124..266P}, and correcting spatial maps of galaxy properties by correcting IFS data cubes for contaminating flux.
Section~\ref{sec:conc} summarizes and conclude the paper.
Appendix~\ref{sec:runningtdose} describes the main routines and setup files of TDOSE and  
Appendix~\ref{sec:TDOSErunexamples} provides a few examples of execution sequences to perform spectral extractions and modify data cubes with TDOSE. 

% ======================================================================
\section{Optimal spectral extraction}\label{sec:optimalext}

% = = = = = = = = = = = = = = = = = = = = = = = = = = = = = = = = = = =
\begin{figure*}
\begin{center}
\includegraphics[width=0.99\textwidth]{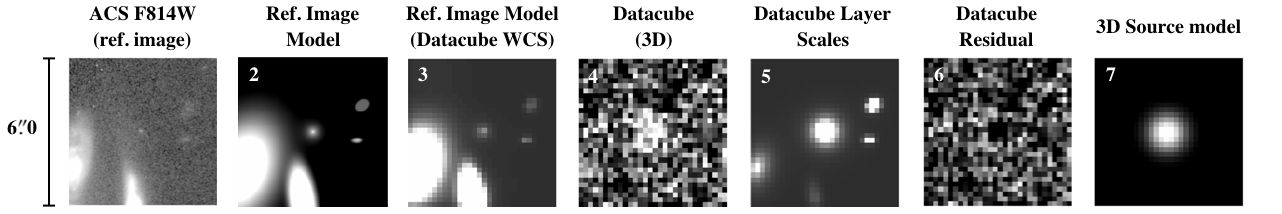}
\caption{Overview of the individual steps and data products produced in the workflow of a standard spectral extraction with TDOSE. 
From left to right, 
1) the HST ACS F814W reference image, 
2) the five-component model of the reference image, 
3) the reference image model converted to the pixel scales of the data cube, 
4) one wavelength layer of the 3D data cube to extract the spectra from, 
5) the flux scaling of the individual model components at the data cube layer and convolved with the data cube PSF, 
6) the residual between the scaled model components and the data cube, and 
7) the 3D source model at the given wavelength layer for the central model component.
Each panel has a size of $6\farcs0\times6\farcs0$ corresponding to a $30\times30$ pixel cutout of the MUSE data cube shown in panel 4. At the chosen wavelength, the central source, a Ly$\alpha$ emitter at $z=3.25$, outshines the other sources in the field-of-view. 
The central model component is scaled accordingly at this particular wavelength when performing the extraction with TDOSE.
Ly$\alpha$ emission is usually \emph{not} well represented by the extent of continuum emission, making a TDOSE extraction based on such a model suboptimal. This is also seen in panel 6, where the extended halo of the Ly$\alpha$ emitter is not recovered by the scaled model, and the central region is over-subtracted.
We caution the use of continuum models for extracting fluxes of emission lines which are poorly represented by the continuum's 2D surface brightness profile.
Instead a more extended "halo" source component can be added to the reference image source model to account for this.}
\label{fig:cubeoverview}
\end{center}
\end{figure*}
% = = = = = = = = = = = = = = = = = = = = = = = = = = = = = = = = = = =

Spectral extraction has been a topic of debate for more than three decades 
since \cite{1986PASP...98..609H} and \cite{1986PASP...98.1220R} formulated a complementary method to standard aperture extractions of slit spectra, that optimizes the S/N in each pixel of the extracted spectrum.
They referred to this as an \emph{optimal} extraction.
With the advent of 3D IFS observations this debate has continued, as an optimal spectral extraction of a 1D spectrum from a 3D IFS data cube should account for the wavelength dependent object morphology, the object's spectral energy distribution variation, any kinematic effects on the spatial distribution of flux as a function of wavelength, as well as the instrumental effects and their variations both spatially and spectrally. 
If this can been done the S/N of the resulting 1D spectrum would also be optimized, i.e. the weighting between actual signal and pixels contributing mostly noise would be accounted for, as was the case for early descriptions of methods to perform optimal extractions from slit-based spectroscopy.

Often spectral extractions are performed with a specific science question in mind, and hence becomes dependent on the science case being investigated. 
For instance, if the goal is to assemble a large sample of spectra for emission line identification and classification, a PSF or "white light" weighted extraction is often enough to obtain the desired results. 
Here "white light" refers to an image obtained by collapsing the IFS data cube along the dispersion direction.
PSF weighted extraction is also useful for studies involving emission line \emph{ratios}, like metallicities and BPT studies, but to get for instance a proper star formation rate estimate correct flux estimates are needed
as much of the line flux could be spatially extended compared to the continuum \citep[e.g.,][]{ForsterSchreiber:2009hm,2016ApJ...828...27N,2016A&A...587A..98W}
And for such estimates, noisy aperture spectra (with or without aperture correction factors) increases the uncertainty in the ability to derive the information from the spectra.
Or if the goal is to estimate emission line equivalent widths of sources with well-detected continuum it is mainly the flux ratios, and the S/N ratio on the measurements which are important. 
Lastly, if the science is focusing on emission expected to deviate from the overall (continuum) morphology of the object, tying the spectral extraction to the object morphology will bias the final results. 
In such cases an optimal extraction will be far from actually being optimal, as the optimized S/N only applies to extractions where the assumed morphology represents the actual data well.

Examples of the latter are studies exploring the spatial extent of Ly$\alpha$ emission, which is known to deviate significantly from the continuum morphology of the host galaxy \citep{2016A&A...587A..98W,2018Natur.562..229W,2017A&A...608A...8L}.
More fundamentally, any nebular emission line, which by definition is not coincident with the stars making up the continuum light distribution, will be biased by a spectral extraction tied to the continuum morphology, if the extent and light distribution of the two are significantly different.
Of course, the significance of such a discrepancy is dependent on the nature of the line, where resonant emission lines, like Ly$\alpha$, must be considered to be the more extreme cases.

Therefore, depending on the science question the extracted spectra are intended to address, alternative methods for spectral extraction might be advisable. 
But generally, for studies where obtaining high S/N of the spectrum is the driving factor, an optimal extraction that accounts for the object morphology and optimizes S/N is preferable.
TDOSE, the tool presented in this paper, provides a broadly applicable software package, for performing such optimal spectral extraction from 3D IFS data cubes.

% ======================================================================
\section{TDOSE}\label{sec:tdose}

TDOSE is a versatile Python software package for extracting one-dimensional spectra and de-blending flux from 3D IFS data cubes.
In this paper we describe version 3.0 of TDOSE \citep{Schmidt:2019cq} but the current "front-end" version of TDOSE is always available from \url{https://github.com/kasperschmidt/TDOSE}.
The main purpose of TDOSE is to optimally extract the flux for individual sources in a given field-of-view (FoV) accounting for both the object morphology as well as the flux from neighboring contaminating sources.
However, as different science cases potentially require different extraction approaches, TDOSE also enables aperture extractions and PSF weighted point source extractions.
Given that all three methods are performed within the same framework and conserve flux, the data products are easily comparable. 

\begin{figure*}
\begin{center}
\includegraphics[width=0.9\textwidth]{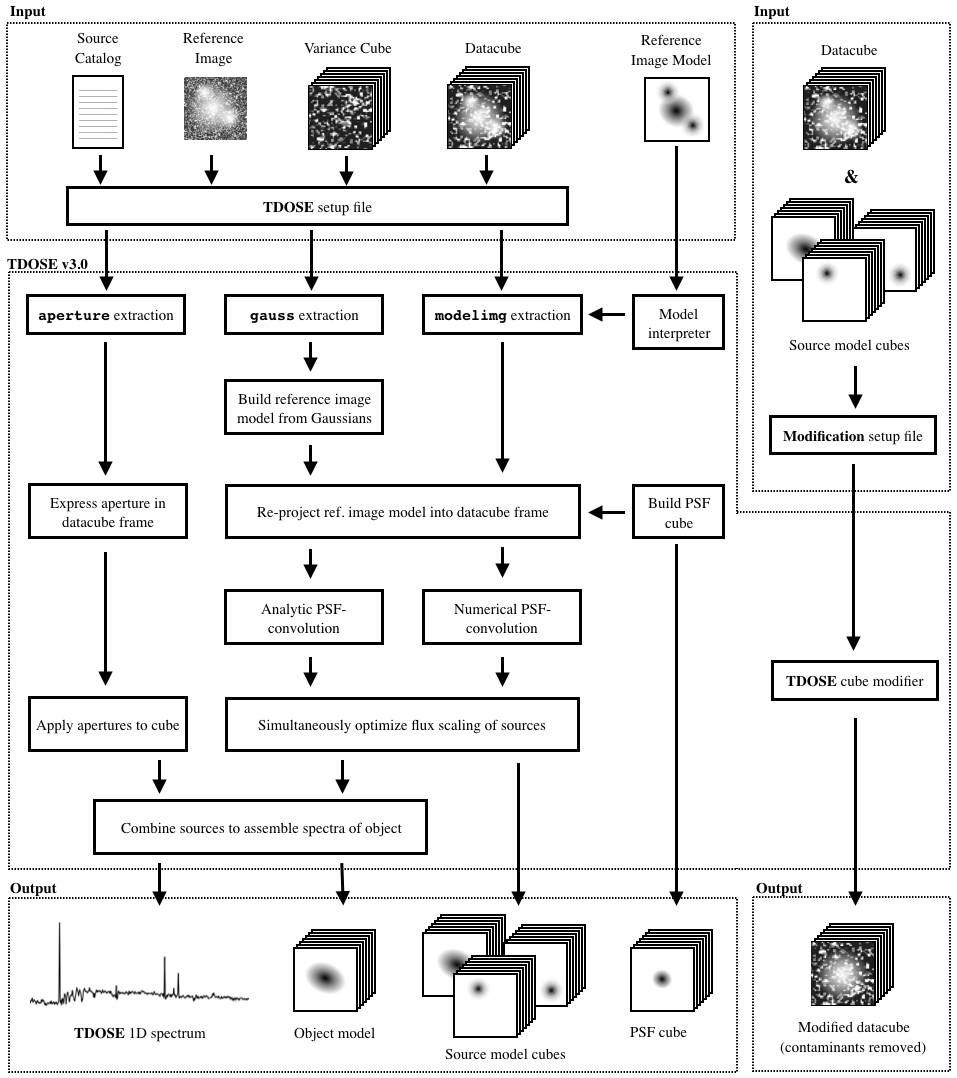}
\caption{Flowchart of TDOSE version 3.0. 
The top layer defines the required (source catalog, reference image, variance cube and data cube) and optional (reference image model and source model cubes) inputs expected by TDOSE for either one of the three extraction methods (\texttt{aperture}, \texttt{gauss} and \texttt{modelimg}; centre left) or the modification of intrinsic data cubes shown on the right.
The bottom part of the panel illustrates the outputs generated by each of the reductions and the cube modifier.
Input reference image models can be converted to the format expected by TDOSE with a "model interpreter". 
Tools for interpreting GALFIT multi-component models and models of individual objects are provided as part of TDOSE.
For details on individual steps, formats and outputs, see Section~\ref{sec:tdose} and Appendix~\ref{sec:runningtdose}.}
\label{fig:flowchart}
\end{center}
\end{figure*}

Figure~\ref{fig:cubeoverview} illustrates the individual steps in the workflow of a standard 3D optimal spectral extraction with TDOSE. 
The workflow can be divided into three main stages:
\begin{enumerate}
\item Determine the sources in the reference image and generate a two-dimensional (2D) morphological model for those (Section~\ref{sec:determinesources}).
\item Convert the reference image model to the IFS reference frame, and determine the flux contribution from each source at each wavelength layer (Section~\ref{sec:fluxscaling}).
\item Combine and de-blend sources in the IFS to extract the 1D spectra of objects in the considered FoV  (Section~\ref{sec:extraction}).
\end{enumerate}
In the following we will describe each of these stages in detail, and explain how aperture and point source spectral extractions are also enabled in the TDOSE software package.
Figure~\ref{fig:flowchart} presents a flow chart of the different spectral extractions, and how IFS data cubes can be modified and corrected for undesirable flux based on the 3D source models generated by TDOSE (cf. Sections~\ref{sec:remsources} and \ref{sec:2Dmaps}).
The TDOSE version 3.0 scripts and setup files used to extract spectra and generate the main outputs,
some of which are displayed in Figure~\ref{fig:cubeoverview} and illustrated in Figure~\ref{fig:flowchart}, are presented in Appendix~\ref{sec:runningtdose}.
Appendix~\ref{sec:TDOSErunexamples} provides examples of a selection of TDOSE tasks and scripts.

The minimum required inputs for TDOSE is a data cube, a variance cube (to propagate and estimate noise on the extracted spectra), a reference image, a model for the PSF wavelength dependence, and a source catalog (see Figure~\ref{fig:flowchart}).
TDOSE is therefore agnostic to the type of IFS data cube the spectra are actually extracted from, as long as the spatial and spectral dimensions are provided in the FITS cube header.
TDOSE was developed with MUSE in mind, and the examples presented in the this paper, are therefore all showing MUSE data and spectra.
Spectral extractions from both CALIFA and MaNGA data cubes have been performed successfully with TDOSE but are not presented in this paper.

% = = = = = = = = = = = = = = = = = = = = = = = = = = = = = = = = = = =
% = = = = = = = = = = = = = = = = = = = = = = = = = = = = = = = = = = =

%-------------------------------------------------------------------------------------------------------------------- 
\subsection{Determining Sources in Reference Image}\label{sec:determinesources}
TDOSE performs optimal spectral extraction and optimizes the S/N by accounting for the spatial morphology and extent of a given object.
Ideally, estimating this morphology would be done on reference imaging with infinite resolution abd a depth exceeding that of the IFS data cube to eliminate any bias from instrumental effects. 
Infinite resolution imaging does not exist but for essentially all existing IFS data, HST imaging or ground-based imaging taken under good conditions can be used.
In Section~\ref{sec:MUSEextractions} we will use HST ACS F814W imaging as reference imaging, when defining the sources that contribute flux to the MUSE data cubes.
Formally, a reference image of the same resolution as the data cube itself, like for instance a white light image, can also be used as the starting point for spectral extractions with TDOSE, if ancillary data is unavailable. 
The use of ancillary (higher-resolution) reference images was deliberately avoided in the spectral extraction tool AUTOSPEC \citep{2018ApJ...869...68G} to make the method self-contained.
However, providing a higher-resolution reference image allows for de-blending of sources which are unresolved at the IFS PSF resolution, which is one of the main strengths of the approach taken by TDOSE.
Avoiding the use of ancillary data could result in extractions of spectra containing flux contamination that could otherwise be avoided. 
Examples of such scenarios are provided in Section~\ref{sec:deblending}. 

After having selected the reference image the sources to base the modeling on have to be determined. 
This can be done using standard imaging source detection softwares like SExtractor \citep{1996A&AS..117..393B}.
Of course, modeling can only be performed on sources.
To be able to account for sources not showing up in the reference image, like emission line sources with faint continuum or sources with abrupt changes in their spectral energy distributions, point sources, or manually generated models, can be used for the spectral extraction. 
Hence, combining standard source detection softwares and source detection tools applied to the IFS data cube itself, like 
the Line Source Detection and Cataloging Tool \citep[LSDCAT;][]{2017A&A...602A.111H}, 
the MUSE Line Emission Tracker \citep[MUSELET;][]{2016ascl.soft11003B}, 
the detectiOn and extRactIon of Galaxy emIssion liNes tool
\citep[ORIGIN;][Mary et al., in prep.]{2017A&A...608A...1B, 2017A&A...608A...2I}, or
the Source Emission Line FInder \citep[SELFI;][]{2016A&A...588A.140M},
can be useful for assembling the most complete source lists and corresponding source models for the spectral extraction.
The important thing to note is that only sources included in the source catalog provided to TDOSE can be accounted for in the spectral extraction.

Having determined which sources to account for in the spectral extraction, a reference image model has to be generated (Figure~\ref{fig:cubeoverview} panel 2). 
This model can either be an empirical representation of the source morphologies in the FoV based on binary or weighted flux segmentation regions similar to the segmentation maps SExtractor produces,
or it can be a sample of analytic parametric 2D light profiles, like \cite{1963BAAA....6...41S} profiles or multivariate Gaussians.
The key point is that each source has a unique representation mimicking its reference image morphology, or rather, the expected underlying morphology of the IFS flux. 

By default, TDOSE models the sources in the reference image by representing each source in the source catalog by a multivariate Gaussian defined as
\begin{equation}\label{eq:multivariategauss}
f(\textbf{c}) = \frac{1}{2\pi\sqrt{\det\boldsymbol{\Sigma}}}\exp\left( -\frac{1}{2}\left(\mathbf{c}-\boldsymbol{\mu}\right)^\textrm{T}\Sigma^{-1} \left(\mathbf{c}-\boldsymbol{\mu}\right)\right)
\end{equation}
where $\mathbf{c}$ represents the coordinate set $(x,y)$ and $\boldsymbol{\mu}$ contains the mean values $(\mu_x,\mu_y)$.
The covariance matrix is given by
\begin{equation}
\boldsymbol{\Sigma} = 
\begin{bmatrix}
    \sigma_x^2 & \rho\, \sigma_x\sigma_y \\
    \rho\, \sigma_x\sigma_y & \sigma_y^2 
  \end{bmatrix}
\end{equation}
where $\rho$ is the correlation between $x$ and $y$.
The morphological multivariate Gaussian models are generated and optimized using Scipy \citep[\url{https://www.scipy.org}; ][]{SciPyOpensources:tUcReTVZ}. 
Should the source list contain objects which are faint or undetectable in the reference image, these are challenging to model automatically with TDOSE. 
In such cases, TDOSE can be instructed to add a point source fixed at each source location to the model.
When to use point sources and when to trust the model depends, among other things, on the completeness of the source catalog and the quality of the reference image.

Alternatively, a custom 2D model of the (reference image) FoV can be provided to TDOSE.
Such a model can for instance be generated with GALFIT models \citep{2010AJ....139.2097P,2002AJ....124..266P}.
TDOSE provides tools to enable de-blending of the individual model components of GALFIT (see Appendix~\ref{sec:TDOSErunexamples}). 
Custom models are treated numerically in the spectral extraction, which in cases of large FoVs increases the computation time and hence the time it takes to extract spectra.

Generating models or providing custom models of the sources in the reference image informs the flux optimization in the second stage of TDOSE about the number and light distribution of the sources to account for during the spectral extraction and source de-bending. 

%-------------------------------------------------------------------------------------------------------------------- 
\subsection{Building a Source Model Cube via Flux Optimization}\label{sec:fluxscaling}

Using the information from the reference image source model, TDOSE optimizes the flux distribution of each wavelength layer, assigning fluxes to each source according to its morphological representation in the reference image model.
The reference image source model is turned into a cube by convolving the reference image model with the wavelength dependent IFS PSF, after pixelating the high-resolution reference image models to the spaxel size of the IFS (Figure~\ref{fig:cubeoverview} panel 3). 
Numerical convolution over large spatial scales at thousands of wavelength layers is computationally expensive. 
TDOSE version 3.0 therefore uses a Gaussian PSF model and by default the multivariate Gaussian source models, such that the PSF convolution is carried out analytically. 
Providing the wavelength dependent IFS PSF as an analytic or empirical function has not been implemented in TDOSE yet.
For source models with non-gaussian source components a direct (non-FFT) numerical convolution with the PSF is performed.
Having transformed the reference image model into a 3D data cube, the flux scalings of the individual source models that best represent the IFS data cube (Figure~\ref{fig:cubeoverview} panel 4) can be determined by solving the set of linear equations defined by 
\begin{equation}
\chi^2 = |\mathbf{A}\mathbf{a}-\mathbf{d}|^2  \; .
\end{equation}
Here, $\mathbf{A}$ represents a list of the 3D models for each of the $n$ sources in the reference image FoV. 
% four-dimensional (4D) matrix representing
The factor $\mathbf{a}$ is a matrix of (flux) scalings for each of the individual source representations in $\mathbf{A}$.
The matrix $\mathbf{d}$ represents the IFS data cube that the source models are supposed to represent.
Hence, given the source models, by minimizing the $\chi^2$ expression, the flux scalings that best represents the IFS data cube can be derived for all sources simultaneously. 
The $\chi^2$ minimization can be done with matrix algebra for each of the $m$ wavelength layers on the IFS data cube using the Ordinary Least Squares (OLS) estimator in matrix form:
\begin{equation}\label{eqn:fluxscales}
\mathbf{a}_m = \left( \mathbf{A}^\textrm{T}_m  \mathbf{A}_m \right)^{-1} \left( \mathbf{A}^\textrm{T}_m \mathbf{d}_m \right) \;.
\end{equation}
Here $ \mathbf{A}_m$ is a matrix of dimension 
$(n, N_\textrm{pix})$ 
representing the source models for each individual wavelength layer. 
The models are normalized by the square root of the variance in each voxel of the wavelength layer. 
The $\mathbf{d}_m$ is a vector of dimension $N_\textrm{pix}$ containing the data flux values normalized by the square root of the variance in each of the $m$ layers. 
Lastly, $\mathbf{a}_m$ is the total flux in layer $m$ assuming the models are normalized (Figure~\ref{fig:cubeoverview} panel 5). 
Hence, $\mathbf{a}_m$ gives the optimized flux, that combined with the given source models, best represents the data in the $m$'th layer of the data cube $\mathbf{d}$ (Figure~\ref{fig:cubeoverview} panel 6).   

This approach is identical to the approach used in PampelMuse \citep{2013A&A...549A..71K,2018ascl.soft05021K} developed to extract and de-blend stellar spectra in crowded IFS data cubes, with the exception that in TDOSE $\mathbf{A}$ consists of extended source models, as opposed to point sources in PampelMuse.

%-------------------------------------------------------------------------------------------------------------------- 
\subsection{Extracting Spectra: De-blending and Combining Sources Into Objects}\label{sec:extraction}

Having minimized the disagreement between the scaled 3D source models and the IFS data cube, the spectrum of any object in the modeled FoV can be extracted.
An object consists of any number, $k$ of the $n$ source model cubes produced by TDOSE.
Any sources that are not included in the object are considered to be contaminants.
Due to the algebraic treatment of the $\chi^2$ minimization described in the previous sub-section, the flux values are optimized simultaneously for all sources in the FoV ensuring ideal conditions for flux de-blending of neighboring sources.
And as the models are based on the (high-resolution) reference imaging it is in principle the resolution of the reference image that drives the ability of TDOSE to de-blend objects from contaminating sources.
Due to the simultaneous flux scaling determined for all sources, the relative flux contribution of each source in the FoV model to each of the IFS voxels is accounted for, and the object flux cube is obtained by simply summing up the flux scaled source model cubes of all the sources contributing to the object (Figure~\ref{fig:cubeoverview} panel 7).
Collapsing this object model cube in the spatial dimensions results in the optimally extracted 1D TDOSE spectrum.
The 1D TDOSE spectrum, the 3D object model and the individual source models can all be returned by TDOSE cf. Figure~\ref{fig:flowchart}.
The latter can be used to modify the original IFS data cube (see Sections~\ref{sec:remsources} and \ref{sec:2Dmaps}).

The noise on the extracted de-blended 1D spectrum is propagated from the variance of the IFS data cube. 
Following Equation~(16) of \cite{2013A&A...549A..71K}, TDOSE estimates the noise at each wavelength in the 1D spectrum as
\begin{equation}\label{eqn:noise}
\textrm{N}_m = \left( \sum_{i,j}  f_{i,j,m}^2 / v_{i,j,m}^2 \right)^{-0.5} \;.
\end{equation}
Here $f_{i,j,m}$ is the fraction of flux in each data cube voxel with respect to the total flux in each wavelength layer $m$ of the data cube, and $v_{i,j,m}$ is the variance of each voxel in the data cube. The sum is performed over the spatial indices $i,j$ of the data cube.
Equation~(\ref{eqn:noise}) results in a S/N at each of the $m$ wavelengths in the extracted 1D spectrum of
\begin{equation}
\left(\frac{\textrm{S}}{\textrm{N}}\right)_m = \frac{\sum_k\mathbf{a}_{m,k}}{\textrm{N}_m} \;,
\end{equation}
where $\sum_k\mathbf{a}_{m,k}$ represents the sum of the flux scales in the $m$'th wavelength layer of the $k$ (out of the total $n$) sources contributing to the object's spectrum.

In the example of the TDOSE extraction workflow presented in Figure~\ref{fig:cubeoverview}, only the source model shown in panel 7 contributes to the object spectrum, i.e. $k=1$. The remaining $n-k=4$ sources seen in the five-component reference image model (panel 2) are considered to be contaminants. 
Section~\ref{sec:multicomp} will present examples of the gain in flux and S/N that can be obtained from MUSE data using multi-source models instead of single-source models.

As mentioned a default spectral extraction with TDOSE is based on an object model consisting of only Gaussian sources combined with a Gaussian PSF model. 
This makes the extraction of spectra fully analytic. 
So-called "multi-Gaussian expansion" (MGE), also known as "Mixture of Gaussian" (MoG) models, have been shown to successfully represent the light distribution of most galaxy types and morphologies in both 1D and 2D \citep[e.g.][]{1992A&A...253..366M,1994A&A...285..723E,1994A&A...285..739E,1995AJ....109..572B,2000ApJ...535..692K,2002MNRAS.333..400C,2013PASP..125..719H,2013MNRAS.432.1894S}.
This makes the 2D version particularly interesting for optimal spectral extraction from IFS data cubes similar to the one performed by TDOSE. 
Representing the PSF by a multi-component MGE model itself adds flexibility and precision to the PSF model without removing the benefits of a fully analytic spectral extraction. 
However, handling of a multi-component MGE PSF model has not yet been implemented in TDOSE.

Simultaneously accounting for and assigning the flux in the IFS data cube to multiple sources in the FoV is exactly what is required for performing reliable de-blending of objects, as it keeps track of the fractional contribution of light in each individual voxel in the IFS data cube from all objects in the FoV.
It is this information that TDOSE uses to de-blend the objects of interest from contaminating sources in the IFS.
The framework that TDOSE and PampelMuse \citep{2013A&A...549A..71K} is based on has previously been shown to effectively handle (point) source de-blending in some of the most crowded fields on the sky, namely globular clusters \citep{2016A&A...588A.148H}.
In a similar manner, TDOSE reliably de-blends extended objects in crowded fields like galaxy clusters or deep extragalactic exposures.
PampelMuse takes advantage of the fact that stars are point sources, and hence are well represented by a scaled PSF at all wavelengths. 
As TDOSE, on the other hand, is designed to handle extended objects and hence requires source modeling, it is worth noticing that the de-blending capabilities of TDOSE are naturally limited by the accuracy of the source model's representation of the actual data, and by the fact that galaxy morphology changes as a function of wavelength, e.g., due to emission line regions/extent and continuum color variations.
As described in Section~\ref{sec:optimalext} a key limitation of any optimal spectral extraction tool using a morphological model as prior, and therefore also TDOSE, is that extractions of spectral features that are poorly represented by the (continuum) morphology described by the reference image model will be biased.
Especially capturing the strength of nebular emission lines which by definition are not described by the continuum light from stars is potentially biased.
To remedy the mismatch between the (continuum) model and intrinsic wavelength dependent 2D morphology when extracting spectra with TDOSE, several approaches can be taken.
For example, the spectral feature, e.g., emission line morphology can be modeled and extracted independently in smaller spectral regions around the observed emission line wavelength, attempting to capture their sizes with dedicated models.
Alternatively, secondary "halo" source components can be added to the intrinsic object model. 
This provides TDOSE with the opportunity to account for emission which extends beyond the compact central continuum emission, by assigning extended emission to a secondary halo source when scaling the model components.
A third alternative, is to estimate the potential bias the continuum-based modeling extraction introduces, and then correct for this.
Section~\ref{sec:deblending} presents a few representative examples of de-blending with TDOSE.

%-------------------------------------------------------------------------------------------------------------------- 
\subsection{Removing Sources from IFS Data Cubes with TDOSE}\label{sec:remsources}

As TDOSE simultaneously models all sources in the FoV, the TDOSE source model cubes can be used for manipulating the original data cube.
One of the main outputs from TDOSE is a flux scaled model of each individual source, which can be collapsed and combined into extracted 1D spectra.
However, instead of collapsing the individual source model cubes, any number of sources can be subtracted directly from the intrinsic data cube.
Hence, TDOSE allows generating data cubes where individual sources (or objects) are removed.
This is useful for science cases where the main goal is to work in 2D or 3D, instead of with extracted 1D spectra. 
Such studies include the description of maps, where satellite galaxies, foreground  and/or background galaxies are contaminating the flux and therefore affecting estimates of the probed parameters of the main galaxy. 
With the optimal de-blending of sources from TDOSE, instead of simply masking the contaminating source positions in for example kinematic, metallicity, emission line or continuum color maps, unmasked full-FoV maps can be generated, after the source models contributing to the contaminating objects are removed from the original IFS data cubes.
This will increase both the areal coverage of the maps, and the precision of the estimates in regions affected by the contaminating sources.
Also, searches for extended low surface brightness emission could potentially benefit from inspection of source-subtracted data cubes. If all sources are removed, the residual data cubes would ideally only contain noise and (extended) emission not captured by the (continuum) source models.
Section~\ref{sec:2Dmaps} presents and example of how to improve a kinematic map of a galaxy by removing contaminating sources.
Appendix~\ref{sec:modcubes} provides an example of the TDOSE commands needed to perform modifications to the original IFS data cube.

%-------------------------------------------------------------------------------------------------------------------- 
\subsection{Point Source and Aperture Extractions}\label{sec:PSFandAperureExtractions}

As a supplement to the optimal spectral extraction, TDOSE also performs point source and aperture extractions. 
A TDOSE point source extraction is simply done by adding a point source to the reference image model, which will then be convolved with the wavelength dependent IFS PSF during the extraction.
This enables extractions using point sources in combination with extended sources.
Such extractions correspond to standard PSF-weighted extractions, except that TDOSE conserves the flux by normalizing the source models before estimating the flux scalings. 
As PampelMuse, which is based on the same framework as TDOSE, is dedicated to spectral extraction in crowded stellar fields like globular clusters, we advise to use this software, as opposed to TDOSE, for such applications.

Aperture extractions with TDOSE are performed by representing each source in the input source catalog by a cylinder in the 3D source model skipping the flux optimization step. 
We note that the apertures are defined based on the (high-resolution) reference image, and are then converted to the IFS voxel scales afterwards, leading to irregular apertures when the IFS's spatial resolution is comparable to the chosen aperture size, i.e., for small apertures with respect to the spatial resolution of the IFS.
Section~\ref{sec:aperturecomp} presents a comparison between aperture and model-based spectral extractions with TDOSE.

% ======================================================================
\section{TDOSE Extractions from MUSE Data Cubes}\label{sec:MUSEextractions}

In the previous section the theoretical framework of the TDOSE software was outlined and described. 
In this section we will illustrate these concepts and the results obtainable by describing and comparing spectra extracted with TDOSE from MUSE IFS data cubes.
In particular, we will focus the comparisons and extractions on objects from the MUSE-Wide survey \citep[see Section~\ref{sec:MW} below and][]{2019A&A...621A.107H,2019A&A...624A.141U}.

%-------------------------------------------------------------------------------------------------------------------- 
\subsection{The MUSE-Wide Survey}\label{sec:MW}

The majority of the data presented in this paper (with the data in Section~\ref{sec:2Dmaps} being the exception), were taken from the MUSE-Wide survey (P.I. L. Wisotzki).
MUSE-Wide is part of the MUSE Guaranteed Time Observations (GTO) campaigns, and comprises 100 MUSE pointings of one square arcminute and 1 hour exposure each mosaiced over the CANDELS \citep{2011ApJS..197...35G,2011ApJS..197...36K} GOODS-South and COSMOS regions.
The first data release from MUSE-Wide (DR1, \url{http://musewide.aip.de}) is described in detail by \cite{2019A&A...624A.141U}, and presents 44 consecutive MUSE pointings collected in CANDELS/GOODS-South.
The 9 pointings of MUSE-Wide covering the Hubble Ultra Deep Field (HUDF) were released (with increased depth) by \cite{2017A&A...608A...1B}.
The remaining 16 fields in GOODS-S, the 2$\times$4 pointings covering the HUDF parallel fields, and the 23 pointings in COSMOS will be released at a later stage.
MUSE-Wide DR1 presents a sample of more than 9000 spectra of photometrically selected objects, including absorption line galaxies, as well as emission line selected galaxies, including more than 400 Ly$\alpha$ emitters (LAEs).
As part of MUSE-Wide DR1, we provided 1D spectra extracted with TDOSE of all objects in GOODS-South from the \cite{2013ApJS..207...24G} source catalog.
Each MUSE-Wide DR1 spectrum was extracted using using the default extraction method of TDOSE version 3.0 (also available in TDOSE version 2.0\footnote{\url{https://github.com/kasperschmidt/TDOSE/releases}.}) described in this paper.
Hence, to represent both the main objects and the contaminating objects for all Guo objects in MUSE-Wide DR1, we used multivariate Gaussian models of the HST F814W image morphology.
To make the extraction process fully analytic, we used Gaussian models for the IFS PSF.
For each MUSE-Wide pointing the wavelength dependent Gaussian PSF is provided in the master PSF catalog selection of MUSE-Wide DR1 \citep[Table~2 of][]{2019A&A...624A.141U}. 
For faint marginally detected objects in the Guo catalogs, where Gaussian modeling was suboptimal, the point source extractions described in Section~\ref{sec:PSFandAperureExtractions} were used.

%-------------------------------------------------------------------------------------------------------------------- 
\subsection{Recovering Spectra of Sources Only Partially Covered}\label{sec:specoutside}

% = = = = = = = = = = = = = = = = = = = = = = = = = = = = = = = = = = =
\begin{figure*}
\begin{center}
\includegraphics[width=0.99\textwidth]{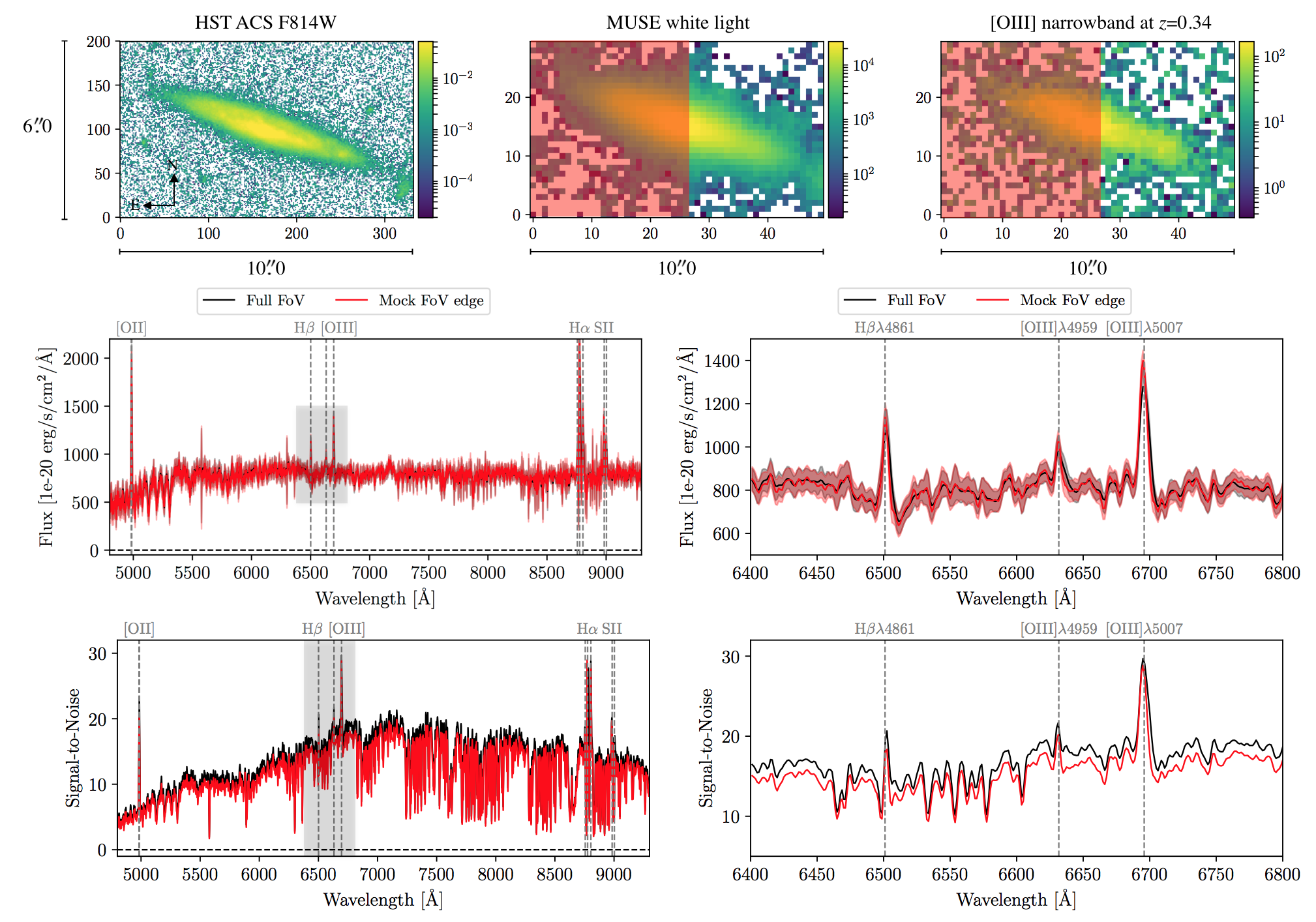}\\
\vspace{-0.4cm}
\caption{Illustration of the recovery of a spectrum from an object (ID$_\textrm{Guo}=10701$, ID$_\textrm{MUSE-Wide}=125034103$) only partially covered (red-shaded region) in the IFS data cube.
From left to right the top panels show $6\farcs0\times10\farcs0$ postage stamps of the HST F814W image, the MUSE white light image and a narrowband image of width 1000km/s (rest-frame) around [OIII]$\lambda$5007\AA. 
The bottom panels show the flux (top) and S/N (bottom) spectra for the full MUSE wavelength range (left) and a zoom-in on the H$\beta$ and [OIII]$\lambda$4959,5007\AA{} emission lines region (marked by the gray box in the left panels).
The black spectrum shows the results from a TDOSE extraction based on a single-source Gauss model of the full FoV shown in the top panels.
The red spectrum on the other hand shows a TDOSE extraction using the same HST source model, but mimicking that the object falls off the edge of the MUSE detector, by only using the area of the MUSE data shaded in red for the model flux scaling in the TDOSE extraction.
The flux levels of the two extractions agrees at the sub-percent level, illustrating the ability to recover the intrinsic flux of an object's spectrum given an assumed morphological model even though (in this case) almost half of the object falls off the IFS detector.
Even though the flux is recovered, the reduced IFS coverage results in a loss of S/N. 
For this example, the median loss in S/N is just below 10\% as illustrated by the S/N spectra in the bottom panels.
}
\label{fig:edge}
\end{center}
\end{figure*}
% = = = = = = = = = = = = = = = = = = = = = = = = = = = = = = = = = = =

As TDOSE bases the spectral extraction on a source model of a high-resolution reference image scaled according to the observed flux in the IFS data cube to obtain the resulting spectrum, the intrinsic object flux is predicted at each wavelength based on the input model. 
This implies that the flux prediction at each wavelength layer of the IFS data cube is insensitive to edges or holes in the data, as it is only based on a scaling of the input model, which is assumed to represent the whole galaxy.
In Figure~\ref{fig:edge} the flux spectrum of ID$_\textrm{Guo}=10701$ (ID$_\textrm{MUSE-Wide}=125034103$) extracted based on the full MUSE data cube (black spectrum) agrees with the flux spectrum extracted from a data cube including a mock edge (red spectrum), where only half of the data (red shaded region in the top panels) were used when determining the flux scalings in Equation~(\ref{eqn:fluxscales}).
The two extracted spectra agree within a median flux difference below one per cent (central panels of Figure~\ref{fig:edge}).
On the other hand, the extracted S/N spectra (lower panels) differ by a median value of 10\% in the shown example. This loss in S/N is caused by the fewer voxels available when extracting the spectrum from only half of the data cube, mimicking that the object falls off the IFS detector.
Hence, assuming that the object model represents the overall source morphology well (in the example shown in Figure~\ref{fig:edge} a single Gaussian model was used), the intrinsic flux can be reliably estimated irrespective of missing data using TDOSE with only a minor loss in S/N.
Multi-component object models, where individual sources fall fully within the excluded (edge) region, will naturally be biased, as those model components cannot be scaled.   
The results will also be biased if only a small fraction of voxels are available or regions that poorly reflect the overall light distribution of the object are used for the flux scaling.

%-------------------------------------------------------------------------------------------------------------------- 
\subsection{Extractions based on Single-Source and Multi-Source Models}\label{sec:multicomp}%\label{sec:MWOIIemitters}

% = = = = = = = = = = = = = = = = = = = = = = = = = = = = = = = = = = =
\begin{figure*}
\begin{center}
\includegraphics[width=0.99\textwidth]{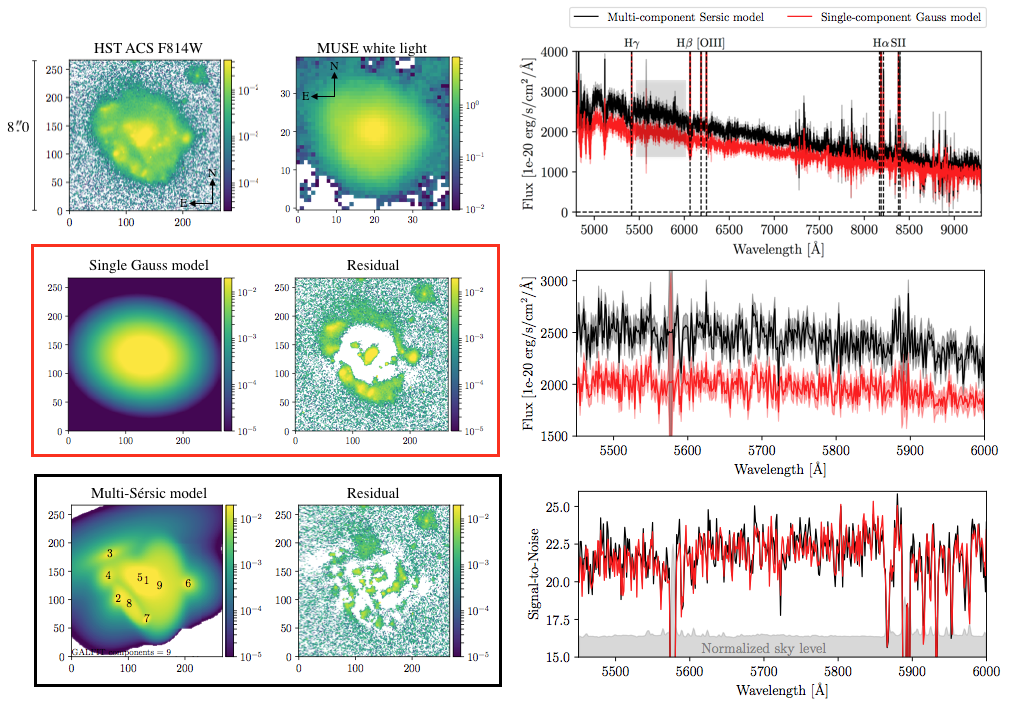}
\caption{Spectral extractions performed with TDOSE for ID$_\textrm{Guo}=10843$ (ID$_\textrm{MUSE-Wide}=112003032$), a star forming H$\alpha\lambda$6563\AA{} emitter at $z=0.2475$.
The lefthand panels show $8\farcs0\times8\farcs0$ postage stamps of the HST F814W image, the collapsed MUSE white light image, a single-component Gaussian model and the corresponding HST residual (red box), and a multi-component S\'ersic model generated with GALFIT and the corresponding HST residual (black box), respectively.
The nine individual components of the GALFIT model are indicated by the black numbers in the bottom left panel. 
The righthand panels show the TDOSE spectra extracted based on the single-component (red) and multi-component models (black) from the MUSE data cube.
The black spectrum combines the flux from all nine sources marked in the bottom left panel. 
The top right panel shows the full spectrum, whereas the bottom right panels show the flux and S/N spectra of the continuum-only region in-between H$\gamma$ and H$\beta$ marked by the gray square in the top right panel.
In the lower right panel the normalized sky flux level responsible for the S/N drops is indicated by the gray shaded region.  
The median flux-increase of $\sim$20\% when using a multi-component model shows the power of basing spectral extractions on multi-component models when the object's light distribution is inhomogeneous.}
\label{fig:10843}
\end{center}
\end{figure*}
% = = = = = = = = = = = = = = = = = = = = = = = = = = = = = = = = = = =

As described in Section~\ref{sec:tdose}, TDOSE is based on a simultaneous scaling a pre-defined sample of sources in the reference image model. 
This model consists of any number of sources that can be combined into spectra of individual objects.
A model is generated by the TDOSE software itself, but can also be provided as an input (see Figure~\ref{fig:flowchart}), and hence be generated manually, or with existing image modeling softwares, like for instance GALFIT  \citep{2010AJ....139.2097P,2002AJ....124..266P}.
The default modeling approach of TDOSE is to assign one multivariate Gaussian (cf. Equation~\ref{eq:multivariategauss}) to each source in the source catalog. 
This allows for both efficient de-blending (Section~\ref{sec:deblending}) and for flexibility to recover the intrinsic flux of non-uniform galaxies as accurately as possible. 
However, as described in Section~\ref{sec:extraction}, depending on the science goal, describing an object by a single source, might not be sufficient.
Figure~\ref{fig:10843} presents an example of the difference between a spectral extraction using a single-source object model and combining multiple source models into a single object.
The figure shows ID$_\textrm{Guo}=10843$ (ID$_\textrm{MUSE-Wide}=112003032$), a star forming galaxy with pronounced features in its 2D light distribution.
The red and black spectra were extracted with TDOSE using the multivariate single Gaussian source model ($n=1$), and the multi-component source model ($n=9$) shown in the bottom left panels.
The models and the image residuals clearly show an improvement in the representation of the 2D light distribution of the galaxy when moving from a single-source (red box and spectra) to a multi-source (black box and spectra) object model.
The recovered continuum flux from the multi-component model is on average 20\% 
higher than the flux level recovered from the single-component model.
The continuum S/N of the two spectra is roughly identical (a median change of 0.5\%) due an on average
higher noise in the spectrum extracted based on the nine-component model compared to the single Gauss extraction.
However, the H$\alpha$ peak S/N increases by roughly 10\% when extracting the spectrum based on the more detailed multi-component model, as the individual star forming regions seen as sub-clumps of the galaxy (and potential differences in the kinematics of these cf. Section~\ref{sec:spatialvar}), are better represented in the multi-component object model.
This illustrates the power of combining models of individual sources, into single objects, when extracting TDOSE spectra for galaxies of inhomogeneous 2D light profiles.

\begin{figure}
\begin{center}
\includegraphics[width=0.5\textwidth]{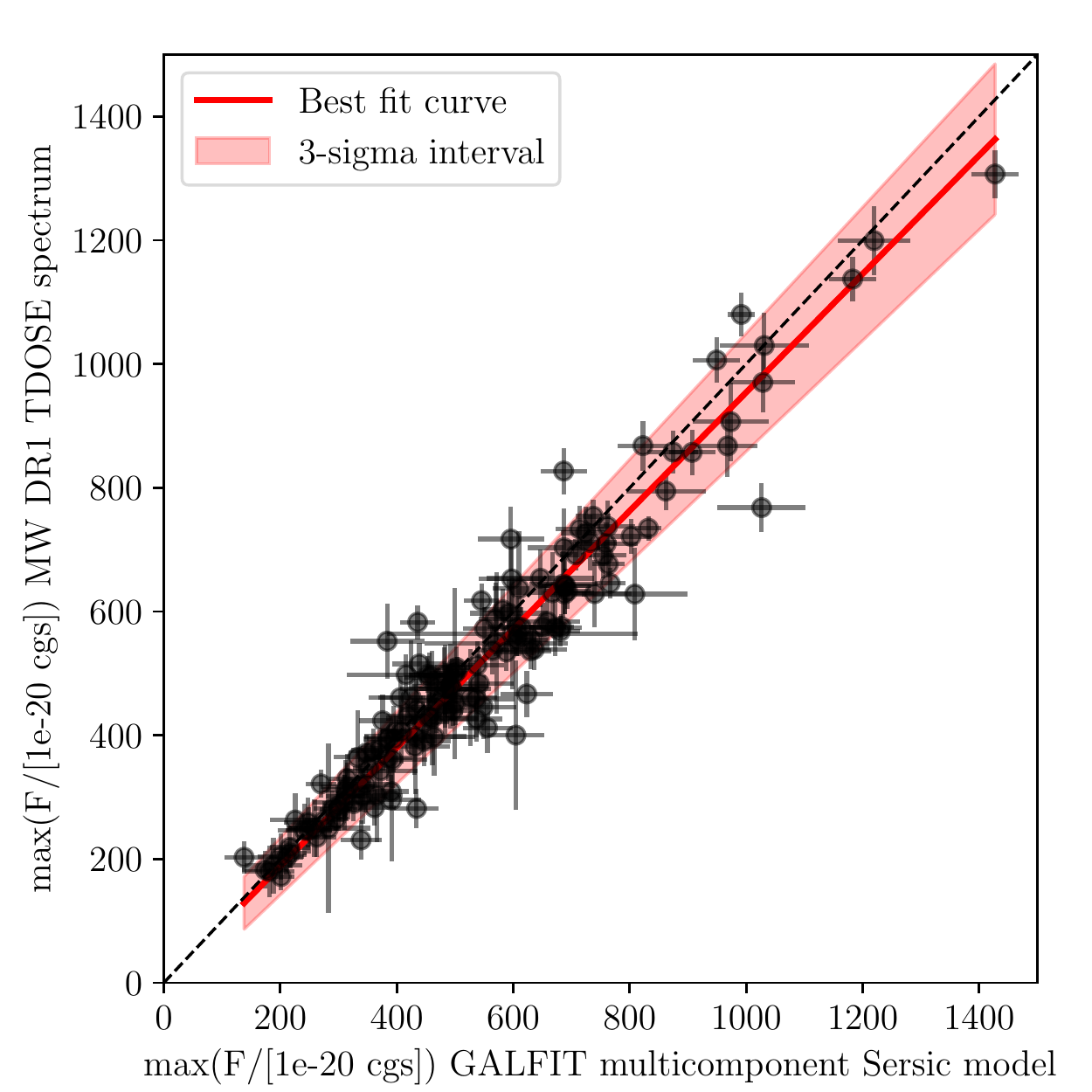}\\
\includegraphics[width=0.5\textwidth]{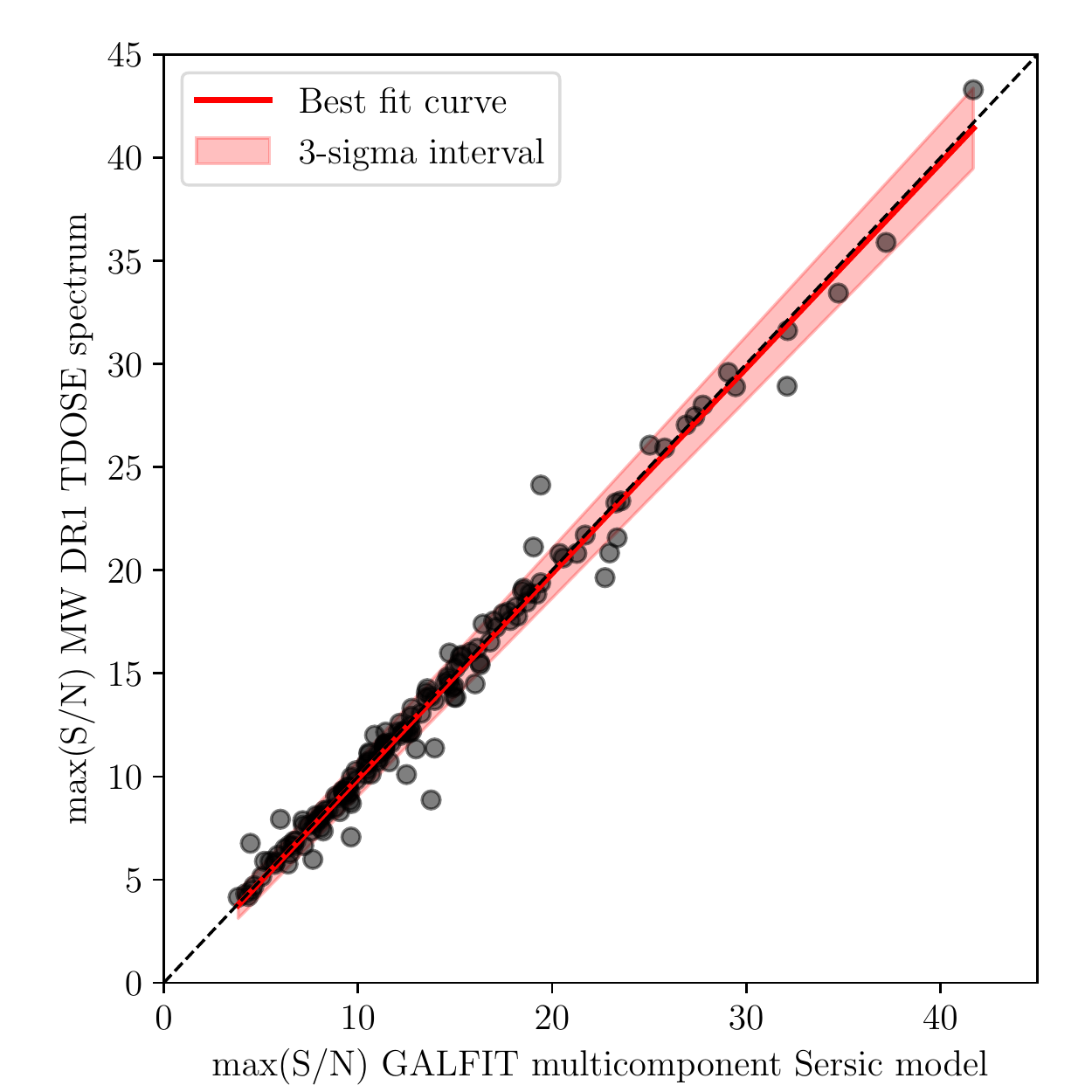}
%\vspace{-0.4cm}
\caption{Comparison of the peak [OII] flux (top panel) and S/N (bottom panel) from the spectra of 153 MUSE-Wide DR1 [OII] emitters extracted with TDOSE based on multi-component S\'ersic models of the HST F814W imaging generated with GALFIT (x-axes), and the MUSE-Wide DR1 TDOSE spectra based on single Gaussian models (y-axes).
In both panels the dashed line marks the one-to-one relation, and the red line is the best linear fit to the flux and S/N measurements with the 3$\sigma$ uncertainty on the fit indicated by the shaded region.
The fit to the flux values in the top panel, which is slightly offset from the one-to-one line, indicates the median 5\% increase in flux gained for this sample when using object models consisting of multiple sources as the base for the spectral extraction.
However, the spread of the data around the best fit shows that the flux increase (or even decrease) gained by using a source model with multiple components can be relatively larger on an object-by-object basis.
The S/N values between the two types of spectra are as shown in the bottom panel in good agreement, despite a few outliers.}
\label{fig:OIIem_TDOSEcomp}
\end{center}
\end{figure}

The object shown in Figure~\ref{fig:10843} is particularly well suited for a multi-source model representation.
To quantify the effects of using multiple versus single source models when extracting spectra of objects in more general terms, we considered a sample of [OII] emitters from MUSE-Wide DR1.
We selected 153 galaxies with apparent magnitudes in the HST F814W filter of $23 < m_{814} < 24$, 
a high-confidence [OII]$\lambda$3726,3729\AA{} emission line doublet identified in the MUSE data cube  \cite["confidence 3" cf.][]{2019A&A...624A.141U},
and a clear match to the photometric \cite{2014ApJS..214...24S} catalog (ID$_\textrm{Skelton} < 0.3$ arcsec).
For this sample, we generated GALFIT multi-source morphological models (with up to 4 independent sources) based on the existing F814W imaging.
We then extracted spectra with TDOSE using both the default single-Gaussian modeling approach (which are the spectra released as part of MUSE-Wide DR1), and by providing multi-source models from GALFIT. 
The top panel of Figure~\ref{fig:OIIem_TDOSEcomp} shows a comparison of the peak flux of the [OII] emission obtained from spectra using the two different source models.
A red line shows the best-fit linear relation between the two flux estimates, obtained from Scipy's \citep{SciPyOpensources:tUcReTVZ} Orthogonal Distance Regression (ODR) in which the uncertainties on both parameters are accounting for.
The median increase in peak flux obtained by using an input model with multiple sources of the object of interest is 5\%, but can be as high as 50\% in extreme cases.
Hence, for sample statistics, the gain in using models containing multiple sources to model and extract spectra for [OII] emitters in MUSE data is relatively modest.
But for individual objects and in special cases the difference can be significant.
Given the extent of the selected [OII] emitters, which is often comparable to the size of the MUSE-Wide PSF ($\lesssim$1'') it is not surprising that the majority of the differences in the HST-based models are eliminated by convolution with the MUSE-Wide PSF.
By comparing the extent of white light, continuum and [OII] emission line narrowband images we confirmed that the [OII] emission is well represented by the (MUSE PSF convolved) continuum models and extent.
Objects that extend well beyond the IFS PSF scales (like ID$_\textrm{Guo}=10843$ shown in Figure~\ref{fig:10843}) will show more significant differences between extractions based on object models with a single and multiple sources. 
The bottom panel of Figure~\ref{fig:OIIem_TDOSEcomp} shows the corresponding S/N for the two spectral extractions. These estimates are in good agreement, with just a few outliers. 
Hence, the main effect of using multiple sources in the object model for the spectral extractions appears to be on the recovered flux levels, as is also confirmed by the extractions for ID$_\textrm{Guo}=10843$.

% = = = = = = = = = = = = = = = = = = = = = = = = = = = = = = = = = = =
% = = = = = = = = = = = = = = = = = = = = = = = = = = = = = = = = = = =

The above examples and comparisons show that the gain in flux between a simple single-source object model and a more complex model including multiple sources, is generally only at the few percent level, but can in special cases be quite high. 
The S/N, on the other hand, appears more stable against variations in the object model.

%-------------------------------------------------------------------------------------------------------------------- 
\subsection{De-blending of objects}\label{sec:deblending}

% = = = = = = = = = = = = = = = = = = = = = = = = = = = = = = = = = = =
\begin{figure*}
\begin{center}
\includegraphics[width=0.95\textwidth]{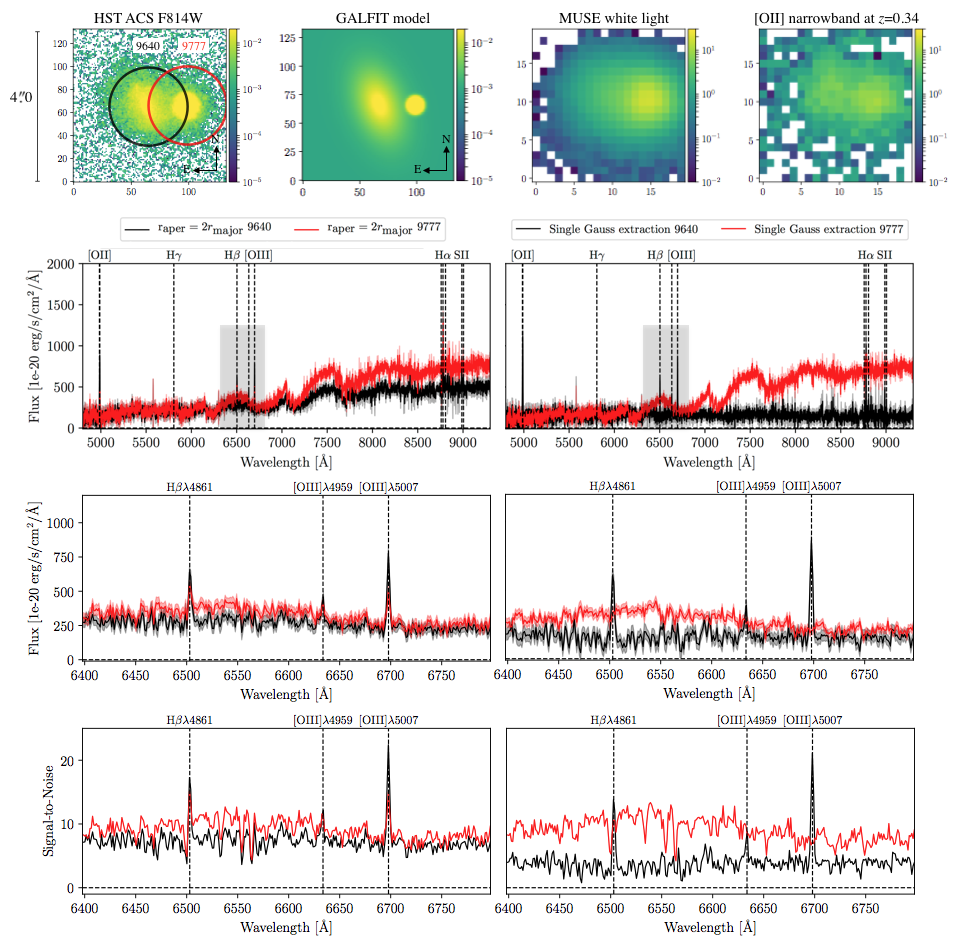}
\vspace{-0.4cm}
\caption{Illustration of extraction of galaxy (object ID$_\textrm{Guo}=9640$, ID$_\textrm{MUSE-Wide}=102009072$) and star (object ID$_\textrm{Guo}=9777$) spectrum while de-blending with TDOSE. 
The top panels show $4\farcs0\times4\farcs0$ postage stamps of the HST F814W image, the two-component GALFIT model, the MUSE white light image and a narrowband around the [OII] emission of the galaxy at $z=0.34$ from the MUSE data cube.
The bottom left panels show the aperture spectra extracted for the two objects, indicated by the circles in the top left panel. The aperture sizes were set to $2\times r_\textrm{major}=1\farcs02$ of the galaxy ID$_\textrm{Guo}=9640$,
providing a good compromise between flux and S/N (cf. Figure~\ref{fig:OIIem_aperturecomp}).
The three panels show the full spectrum and zoom-ins on the H$\beta$ and [OIII]$\lambda$4959,5007\AA{} emission lines region (marked by the gray box in the top panel).
It is clear from these spectra that the galaxy (black) is heavily contaminated by the stellar continuum, and the spectrum of the star (red) shows emission line flux spilling over from the galaxy.  
The bottom right panels are identical to the bottom left panels, except that now the spectra shown were extracted with TDOSE based on the two-component GALFIT model (top second panel).
Through TDOSE's de-blending, the galaxy emission line flux is now completely gone from the stellar spectrum (red) while the continuum flux and S/N are both conserved.
Simultaneously, the continuum level in the galaxy spectrum (black) is much less affected by the neighboring bright star, while the TDOSE spectrum still provides high S/N in the emission lines.}
\label{fig:specs9640} % 9640 <-> 102009072
\end{center}
\end{figure*}
% = = = = = = = = = = = = = = = = = = = = = = = = = = = = = = = = = = =

Apart from improving the flux recovery of individual sources, extracting spectra based on models containing multiple sources can be used to efficiently de-blend spectra from independent objects that appear close to each-other when projected on the sky. 
For the case of MUSE data taken without adaptive optics (AO), HST reference images generally have a resolution that is at least five times better, as the PSF FWHM of HST/ACS images is generally below 0$\farcs$1 \citep{2013ApJS..207...24G}.
The MUSE-Wide non-AO data were in seeing between 0$\farcs$7 and 1$\farcs$2 \citep{2019A&A...624A.141U}.
This implies that it is straight forward to distinguish and reliably model objects which are blended in the MUSE data.

% = = = = = = = = = = = = = = = = = = = = = = = = = = = = = = = = = = =
\begin{figure*}
\begin{center}
\includegraphics[width=0.93\textwidth]{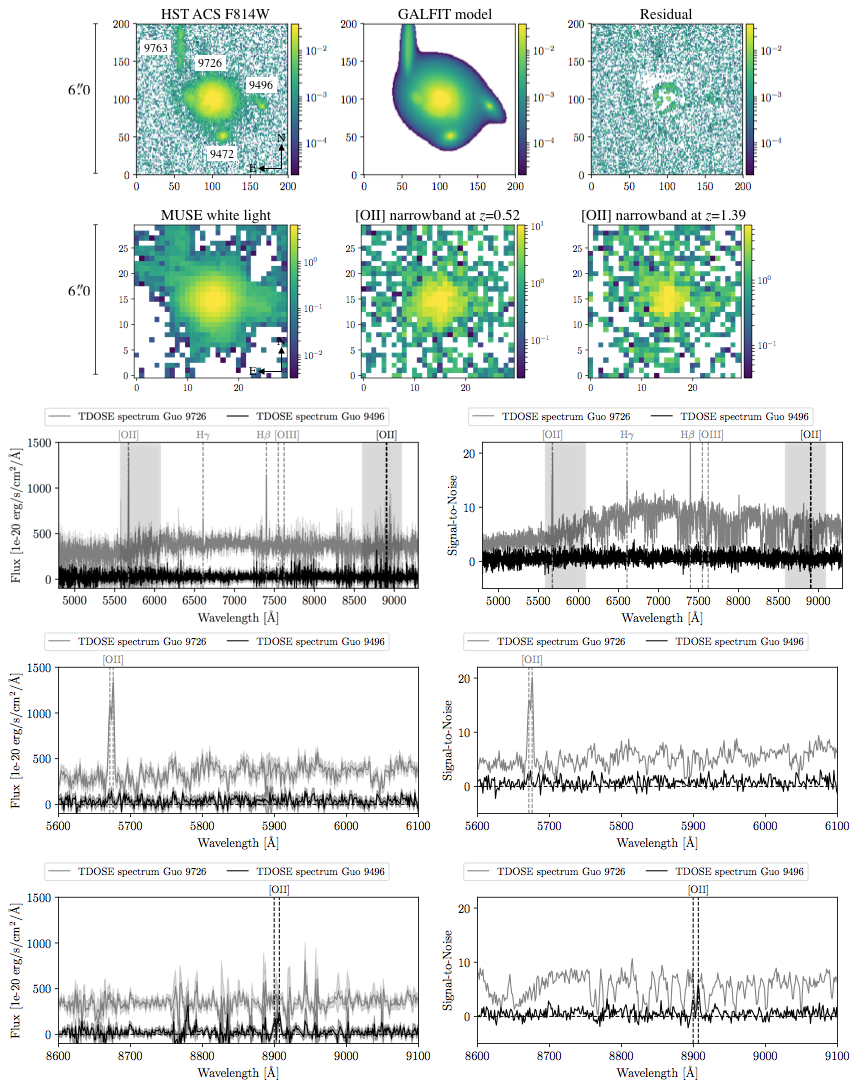}\\
\caption{Illustration of extraction of multiple spectra while de-blending with TDOSE. 
The top panels show $6\farcs0\times6\farcs0$ postage stamps of the (from top left to bottom right) HST F814W image, the multi-component GALFIT model and residual, the MUSE white light image and two narrowbands around [OII] of widths 1000km/s in rest-frame positioned at redshifts 0.52 and 1.39, corresponding to the redshifts of object ID$_\textrm{Guo}=9726$ (ID$_\textrm{MUSE-Wide}=125017033$) and ID$_\textrm{Guo}=9496$ (ID$_\textrm{MUSE-Wide}=125068147$), respectively.
In the bottom panels, the TDOSE spectra extracted for object ID$_\textrm{Guo}=9726$ (gray) and ID$_\textrm{Guo}=9496$ (black) based on the multi-component GALFIT model (top central panel) are shown. 
The first two panels show the full spectra and the position of prominent emission lines at the redshift of the two objects. The gray shaded regions mark the location of the flux and S/N zoom-ins shown in the bottom two panels.
Due to the efficient de-blending by TDOSE, the continuum and emission line contamination in the spectra is low, and reliable flux estimates can be obtained based on these extractions.
Figure~\ref{fig:9726apercomp} shows a comparison between aperture spectra and the model-based TDOSE spectrum of ID$_\textrm{Guo}=9726$.
}
\label{fig:9726deblend}
\end{center}
\end{figure*}
% = = = = = = = = = = = = = = = = = = = = = = = = = = = = = = = = = = =

Figure~\ref{fig:specs9640} shows images and spectra of the object ID$_\textrm{Guo}=9640$ (ID$_\textrm{MUSE-Wide}=102009072$) at redshift $z=0.3377$ which has strong [OII]$\lambda$3726,3729\AA, H$\beta$, [OIII]$\lambda$4959,5007\AA{} and H$\alpha$ emission in the MUSE-Wide DR1 data cube.
Projected on the sky, the object appears close to the foreground M-star ID$_\textrm{Guo}=9777$.
The two objects are separated in the HST data (top left panel) but are marginally resolved in the MUSE data cube as illustrated in the white light and [OII] narrowband postage stamps in the top right panels given the MUSE PSF size of just below 1$\farcs$0.
The [OII] narrowband has a rest-frame width of 1000km/s.
The amount of contamination, i.e., blending between the spectra of the two objects, strongly depends on the spectral extraction method.
The bottom left panels show aperture spectra for the two objects extracted using an aperture size of $2\times r_\textrm{major}=1\farcs02$, where $r_\textrm{major}$ is the isophotal major axis of ID$_\textrm{Guo}=9640$ provided in the photometric \cite{2013ApJS..207...24G} source catalog.
As we will show in Section~\ref{sec:aperturecomp} $2\times r_\textrm{major}$ provides a good compromise between recovered flux and optimal S/N for aperture extractions, where you cannot obtain both.
As the FWHM of the MUSE PSF is $\sim1\farcs0$ this aperture size also recovers the vast majority of the light from the star. 
The characteristic "wavy" continuum of the M-star (red spectrum) is clearly seen imprinted on the extracted galaxy spectrum (black spectrum). And in a similar way, the stellar spectrum has been contaminated by the galaxy emission lines.
Likewise, a MUSE PSF-weighted spectral extraction results in spectra with considerable contamination.
The spectra extracted with TDOSE shown in the bottom right panels are based on a single-source model for each of the two objects (generated with GALFIT and shown in the top second panel), and are significantly cleaner than the aperture (and PSF weighted) spectra.
Due to the efficient de-blending by TDOSE the galaxy now has a flat low-level continuum without obvious imprints from the neighboring star, and the stellar spectrum has been cleaned for the galaxy emission lines.
The de-blending and spectral extraction by TDOSE was done without any loss in flux or S/N compared to the $2\times r_\textrm{major}$ aperture extractions.  

Figure~\ref{fig:9726deblend} shows a second example of de-blending with TDOSE from solving the equations presented in Section~\ref{sec:fluxscaling}, 
The complex of four individual Guo objects (top left panel) were modeled using multiple sources with GALFIT (top central and right panel).
This reference image model was used to extract spectra from the MUSE-Wide datacube where the flux is blended, as illustrated by the MUSE white light image also shown.
The bottom panels show the spectra for ID$_\textrm{Guo}=9726$ (ID$_\textrm{MUSE-Wide}=125017033$, gray spectra) and ID$_\textrm{Guo}=9496$ (ID$_\textrm{MUSE-Wide}=125068147$, black spectra).
The bright continuum of ID$_\textrm{Guo}=9726$ has been cleanly de-blended from the fainter line emitter ID$_\textrm{Guo}=9496$. 
Also, the cross-contamination from the [OII] emission in the two objects which are at $z=0.52$ and $z=1.39$, respectively, has been removed with the de-blending by TDOSE.

Another case of de-blending happens when objects in the reference (HST) imaging are only barely resolved, but the IFS data show clear spectral features of an un-resolved superposition of multiple sources. 
In such cases the centroid of the spectral features in the IFS data cube can be used to de-blend and assign flux to independent sources of one (or more) objects in the FoV.
Such de-blending can determine the origin of for instance emission lines and other prominent spectral features, and through this, reliably determine redshifts of objects that are only marginally resolved at the resolution of the reference imaging used.
In such cases, photometric catalogs often only assign a single ID to the unresolved objects.
To avoid too aggressive photometric de-blending this is likely also the correct thing to do, when the de-blending based on the IFS data is unavailable.
However, if physical parameters like stellar mass, equivalent width, or SFR, are estimated from fitting templates to the photometry of the combined flux the results would be biased, given that the combined flux from the unresolved objects is assumed to be from a single source with the spectral features (e.g., emission lines) from the IFS data cube.

We note that the simultaneous modeling and de-bending of objects with TDOSE introduces non-zero covariances between the different source models. These covariances can become significant for especially unresolved sources.  
Nevertheless, the above examples illustrate the importance of careful spectral extraction, using the best possible reference imaging, to avoid mis-identification and interpretations, especially at high redshift where blending is very prominent in most IFS data.
If the use of ancillary data had deliberately been avoided in the examples shown in this section, the scientific analysis of the extracted spectra and the corresponding broad band photometry would have been biased.

%-------------------------------------------------------------------------------------------------------------------- 
\subsection{Comparison of Model-based and Aperture-based Extractions}\label{sec:aperturecomp}

% = = = = = = = = = = = = = = = = = = = = = = = = = = = = = = = = = = =
\begin{figure}
\begin{center}
\includegraphics[width=0.49\textwidth]{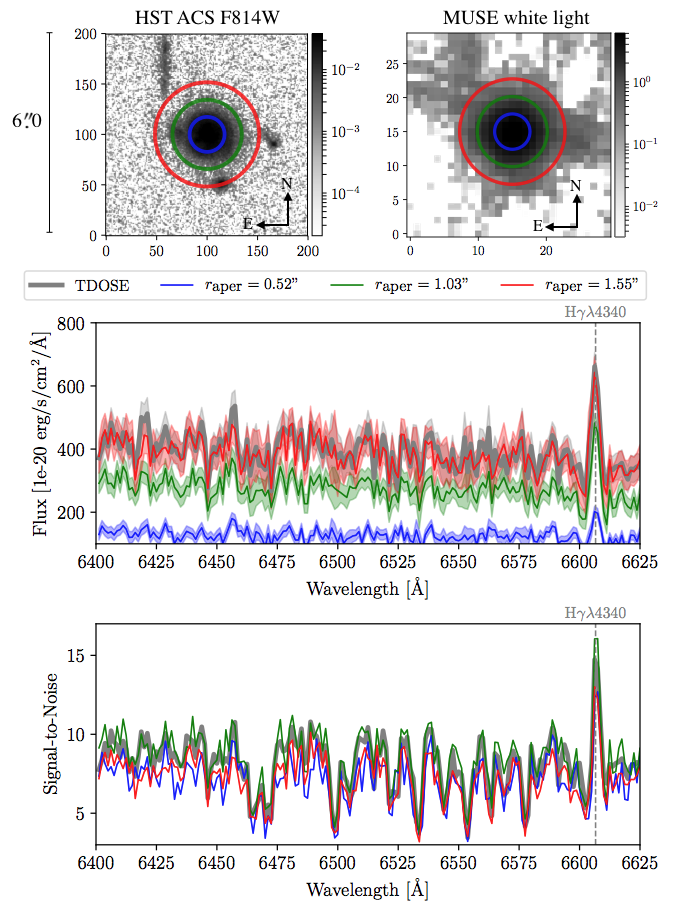}\\
\caption{Comparison of three different aperture spectra (blue, green and red) with the TDOSE spectrum of ID$_\textrm{Guo}=9726$ shown in Figure~\ref{fig:9726deblend} (thick gray) which was extracted based on a multi-component object model. 
The radii of the apertures are $1\times r_\textrm{major} = 0\farcs52$ (blue), $2\times r_\textrm{major}=1\farcs03$ (green) and $3\times r_\textrm{major}=1\farcs55$ (red).
The top panels show $6\farcs0\times6\farcs0$ postage stamps of the HST F814W and the MUSE white light images with the aperture sizes marked by the colored circles.
The bottom panels show a zoom-in on the emission line free continuum blue-wards of the H$\gamma$ line.
The largest aperture (red) recovers most flux but suffers from a lower S/N. 
On the other hand the median-aperture extraction (green) provides the highest S/N but is suffering from a lower recovered flux. 
The spectrum extracted with TDOSE (thick gray) recovers the same flux as the large-aperture spectrum but \emph{also} has a S/N similar to the median-aperture extraction.
Hence, this illustrates that TDOSE provides \emph{both} high flux and high S/N as opposed to simple aperture extractions.
Figure~\ref{fig:OIIem_aperturecomp} provides similar conclusions for a sample of 153 fainter [OII] emitters.
}
\label{fig:9726apercomp}
\end{center}
\end{figure}
% = = = = = = = = = = = = = = = = = = = = = = = = = = = = = = = = = = =

As described in Section~\ref{sec:optimalext} an optimal spectral extraction attempts to recover the intrinsic flux of the considered object as accurately as possible, while still providing a high S/N by limiting the amount of voxels with limited information included in the extraction.
On the other hand, standard aperture extractions implicitly requires a choice between accurate flux measurement or high S/N. An aperture extraction cannot provide both.
This is illustrated in Figure~\ref{fig:9726apercomp}, where we show the emission line free region of the continuum bluewards of the H$\gamma$ line for the object ID$_\textrm{Guo}=9496$ also shown in Figure~\ref{fig:9726deblend}.
The three apertures marked by the colored circles in the top panels, have a size of $1\times r_\textrm{major} = 0\farcs52$ (blue), $2\times r_\textrm{major}=1\farcs03$ (green) and $3\times r_\textrm{major}=1\farcs55$ (red).
Here $r_\textrm{major}$ corresponds to the isophotal major axis (\verb+a_image+) of ID$_\textrm{Guo}=9496$ as  measured by SExtractor and provided in the \cite{2013ApJS..207...24G} catalog. 
The bottom panels show that the largest aperture as expected captures the largest amount of flux at cost of a lower overall S/N (bottom panel). 
A good compromize between flux recovery and S/N appears to be the extraction using an aperture with a radius of $2\times r_\textrm{major}$ (green circle and spectra).
The thick gray curve shows the TDOSE spectrum of ID$_\textrm{Guo}=9496$ also presented in  Figure~\ref{fig:9726apercomp}.
The TDOSE spectrum recovers the flux at the level of the $3\times r_\textrm{major}$ aperture spectrum (red, central panel), but \emph{simultaneously} provides a S/N per pixel comparable to the $2\times r_\textrm{major}$ aperture spectrum (green, bottom panel).

\begin{figure}
\begin{center}
\includegraphics[width=0.5\textwidth]{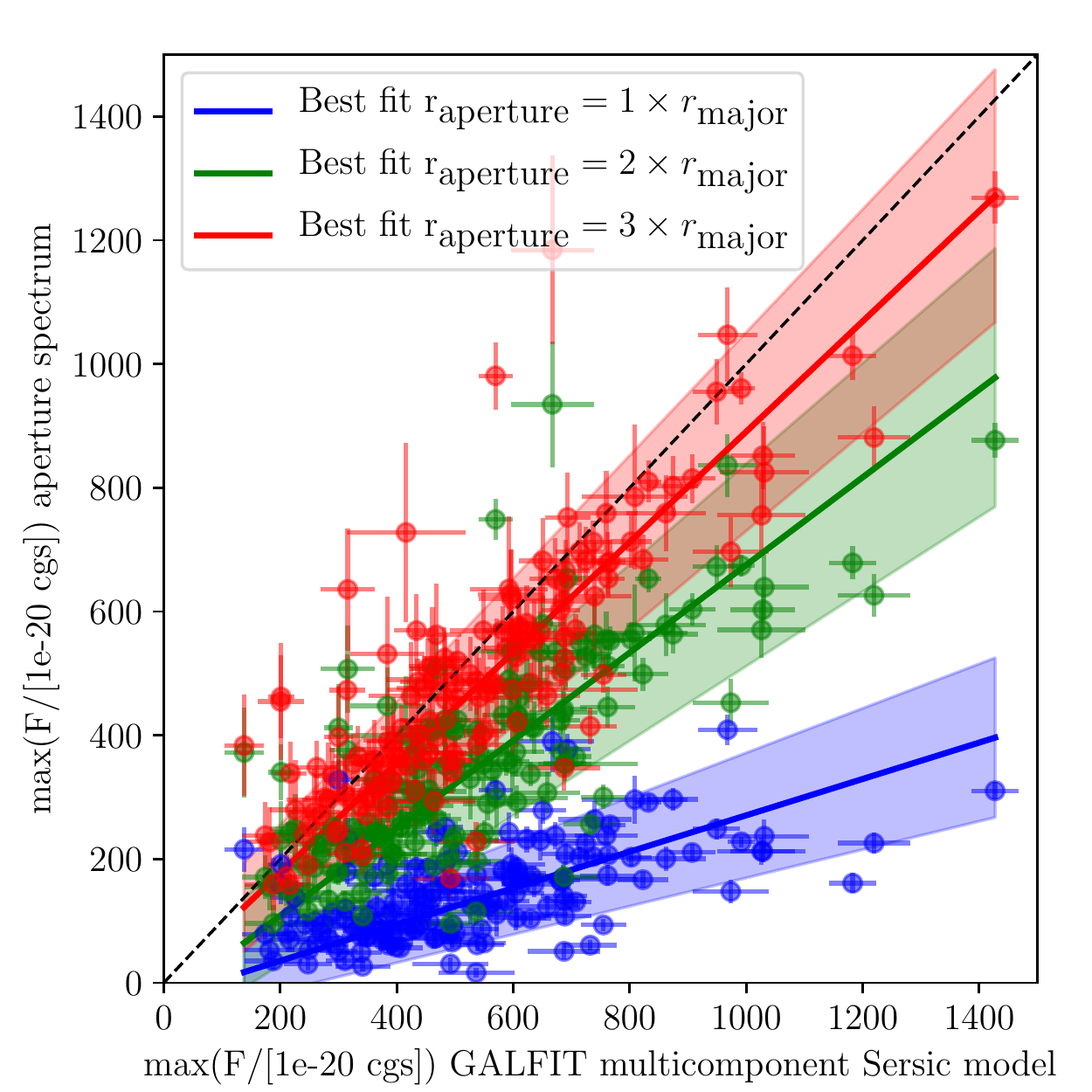}\\
\includegraphics[width=0.5\textwidth]{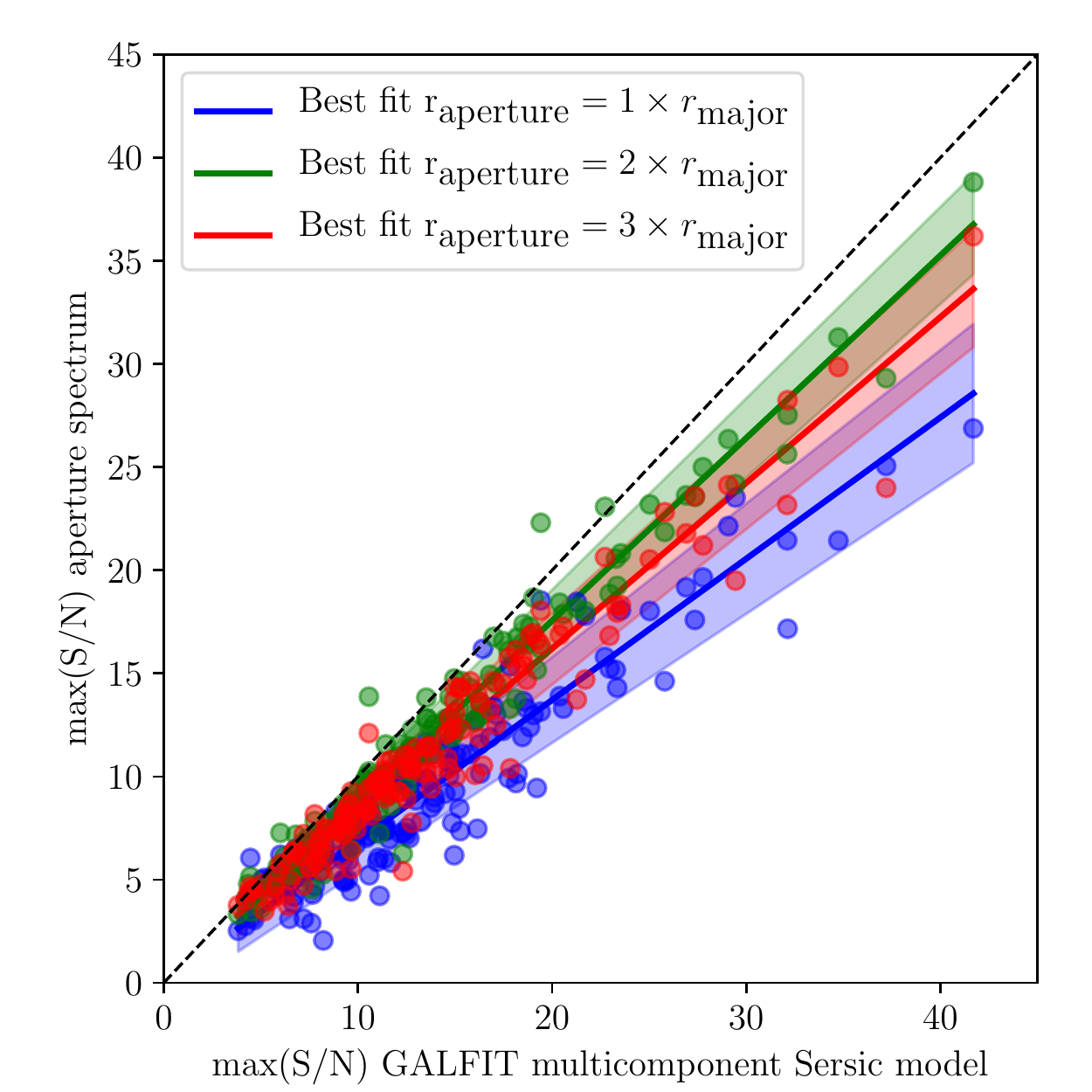}
\caption{Comparison of the peak [OII] flux (top panel) and S/N (bottom panel) from the spectra of the 153 MUSE-Wide DR1 [OII] emitters also analyzed in Figure~\ref{fig:OIIem_TDOSEcomp} extracted with TDOSE based on multi-component S\'ersic models of the HST F814W reference imaging generated with GALFIT (x-axes), and apertures with radii equal to 1, 2 and 3 times the major axis of each object provided in the photometric catalogs by \cite{2014ApJS..214...24S} plotted in blue, green and red, respectively.
In both panels the dashed line marks the one-to-one relation, and the solid lines indicate the best linear fits to the individual flux and S/N measurements with the 3$\sigma$ uncertainty on each fit indicated by the shaded regions.
In all cases the average flux and S/N is higher for the model-based TDOSE extractions, even though the median flux increase is only $\sim$10\% for the largest flux aperture.
Hence the TDOSE extractions based on GALFIT models avoid the compromise between flux and S/N necessary for aperture extractions, illustrated by the shift of the order between the three aperture samples, and accurately recovers the flux while still providing high S/N.}
\label{fig:OIIem_aperturecomp}
\end{center}
\end{figure}

To estimate the efficiency of the TDOSE extractions compared to aperture extractions for a larger sample of objects, we return to the $\sim$150 MUSE-Wide [OII] emitters considered in Section~\ref{sec:multicomp}.
We extracted aperture spectra for all objects again using aperture radii of $1\times r_\textrm{major}$ (blue), $2\times r_\textrm{major}$ (green) and $3\times r_\textrm{major}$ (red).
The $r_\textrm{major}$ value for each object was taken from the \cite{2014ApJS..214...24S} catalog.
In Figure~\ref{fig:OIIem_aperturecomp} we compare the peak flux (top) and S/N (bottom) of the [OII] emission line from the TDOSE extractions based on GALFIT multi-component models (x-axis) and the aperture spectra (y-axes).
As expected, the largest apertures on average recovers the most flux, whereas a more modest aperture of just $2\times r_\textrm{major}$ achieves the highest S/N on average in agreement with the extractions for ID$_\textrm{Guo}=9496$ in Figure~\ref{fig:9726apercomp}.
As was the case for ID$_\textrm{Guo}=9496$ the model-based TDOSE extractions are capable of providing a high S/N while still delivering a reliable estimate of the peak [OII] flux for the MUSE-Wide [OII] emitter sample.
The median increase in flux (S/N) of the TDOSE spectra is even 9\% (14\%) when compared to the $3\times r_\textrm{major}$ ($2\times r_\textrm{major}$) aperture extractions.

Hence, the model-based spectral extractions from TDOSE provide optimal extractions bringing the "best of two worlds" by optimizing S/N while still recovering a large fraction of the emitted flux.

% = = = = = = = = = = = = = = = = = = = = = = = = = = = = = = = = = = =
% = = = = = = = = = = = = = = = = = = = = = = = = = = = = = = = = = = =

%-------------------------------------------------------------------------------------------------------------------- 
\subsection{Generating Contamination Free 2D Maps}\label{sec:2Dmaps}

As described in Section~\ref{sec:remsources} the 3D source models produced by TDOSE are useful for removing  unwanted contaminating flux to generate new corrected data cubes. 
This process is illustrated in the right-hand side of the TDOSE version 3.0 flowchart in Figure~\ref{fig:flowchart}.
As mentioned, such modifications of the intrinsic IFS data cubes can improve the S/N, extent and reliability of spatial maps derived from the data cubes.

\begin{figure*}
\begin{center}
\includegraphics[width=0.75\textwidth]{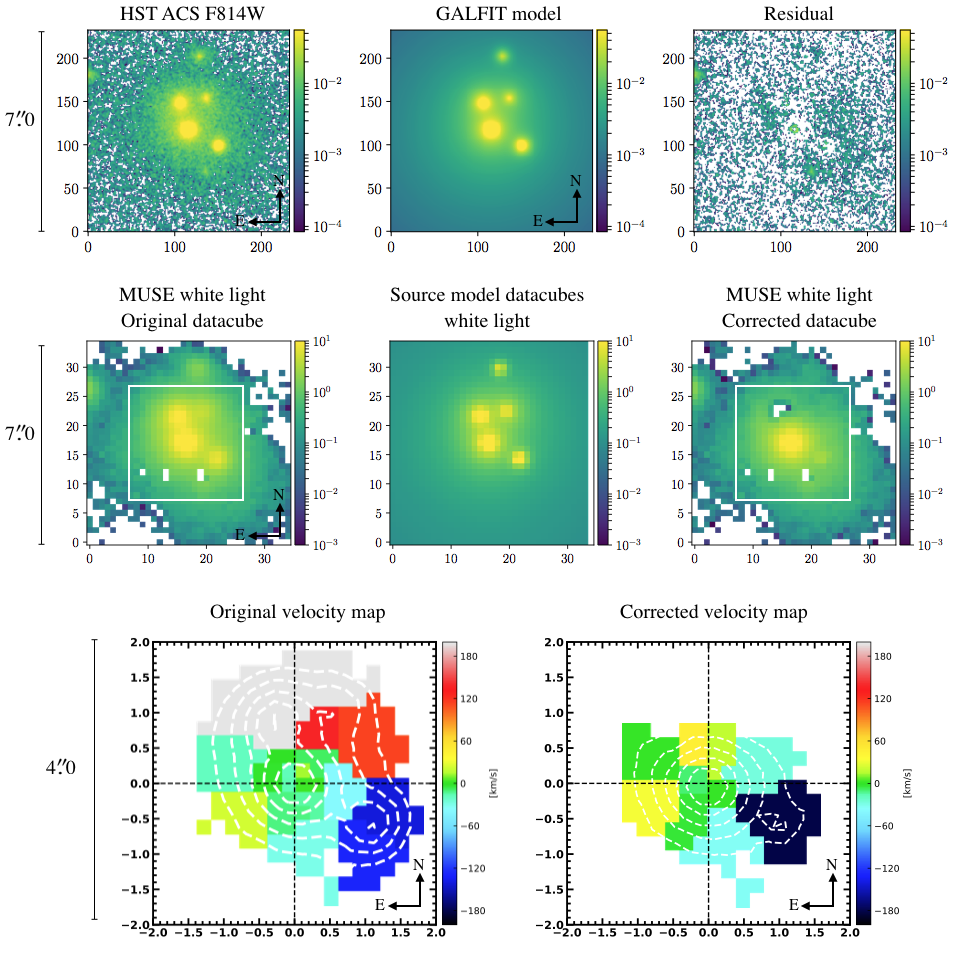}
\caption{Illustration of the improved velocity map obtained after removing the 3D source models of the contaminants generated by TDOSE from the original IFS data cube.
The top panels show the $7\farcs0\times7\farcs0$ postage stamp of the HST F814W image, the HST multi-component GALFIT model, and the corresponding residual image.
The central panels show the $7\farcs0\times7\farcs0$ postage stamp of the MUSE white light image, the HST model in the MUSE coordinate system, and the MUSE white light image after removing all but the central 3D source model generated by TDOSE.
The bottom left panel shows the velocity map generated from the intrinsic MUSE data cube (see Section~\ref{sec:2Dmaps} for details).
Due to the contaminating flux, the velocity estimates might lead to concluding that the central galaxy is rotating.
The bottom right panel shows a velocity map constructed the same way, but now based on the MUSE cube after the contamination has been removed using the TDOSE source models. 
As opposed to the un-corrected velocity map, this version indicates that the central galaxy has no significant rotation.
The extent of the velocity maps are marked by the white squares in the MUSE white light images.
We note that the Voronoi bins in the two velocity maps, as well as their extent, are different as we are left with the flux belonging only to the galaxy of interest after the contaminating flux was removed (bottom right panel).
Hence, removing the contaminants from the IFS data cube, provide an alternative, but more reliable assessment of the 2D velocity map.
}
\label{fig:mapcorrection}
\end{center}
\end{figure*}

To illustrate this use of the TDOSE output we focus on the objects shown in Figure~\ref{fig:mapcorrection} from the galaxy group COSMOS-Gr32 ($z = 0.73$) presented by \cite{2012ApJ...753..121K} and Boselli et al., in prep..
These objects were observed for 5.25 hours with MUSE as part of the GTO program focusing on the effect of environment on galaxy evolution processes (PI: T. Contini).
At this depth the S/N is high enough to derive kinematic maps for the central galaxy in the HST ACS F814W image shown in the top left panel of Figure~\ref{fig:mapcorrection}.
In the $7\farcs0\times7\farcs0$ postage stamps four sources are contaminating the signal obtainable from the central object. 
By modeling all sources in the FoV with GALFIT (top central panel of Figure~\ref{fig:mapcorrection}) and generating the 3D source models with TDOSE, the flux from the contaminants can be removed from the original MUSE data cube as illustrated by the original and contamination-corrected MUSE white light images shown in the central panels of Figure~\ref{fig:mapcorrection}.
We derived the stellar velocity maps for both the original and the contamination-corrected MUSE data cubes using the pPXF method \citep{2004PASP..116..138C}.
To ensure reliable fits to the data, we used Voronoi binning \citep{2003MNRAS.342..345C} requiring that each bin had S/N$>$10.
The voxels in each Voronoi bin were collapsed into a single 1D spectrum and used to estimate the velocities. 
The resulting velocity maps are shown in the bottom panels of Figure~\ref{fig:mapcorrection}.
Without correcting the data cube for contaminating flux the velocity map (bottom left panel) indicates rotation of the central galaxy around an approximate E-W axis, even though such a conclusion is naturally uncertain given the contamination. 
Instead, as the velocity field is not that of the central galaxy, but rather that of the system as a whole, we likely see the relative motions of the contaminating sources, which are all at the group redshift of $z=0.73$.

After removing the contaminating flux, the velocity map is significantly cleaner.
The dark Voronoi bin to the sourth-west still shows signs of contamination residuals. 
These potentially results from kinematic signatures in this galaxy that are poorly represented by the single-Gaussian model (cf. Section~\ref{sec:spatialvar}).
Nevertheless, from the cleaned velocity map (bottom right panel of Figure~\ref{fig:mapcorrection}) it becomes clear that the central object likely has no significant rotation, and if any, the rotation appears to be around a SE-NW axis and not the E-W axis implied by the original map.
Hence, by using the TDOSE source models to correct the original data cube, the conclusions drawn from the resulting velocity map are significantly changed and are more reliable.

Improvements similar to these in maps of physical parameters are obviously not restricted to kinematic maps, as the velocity maps chosen as an example here.
Source models like the ones produced by TDOSE can be used to modify and correct maps of metallicity, star formation rate, electron density etc. generated from IFS data cubes where 3D flux contamination has been corrected for.

% = = = = = = = = = = = = = = = = = = = = = = = = = = = = = = = = = = =
% = = = = = = = = = = = = = = = = = = = = = = = = = = = = = = = = = = =

%-------------------------------------------------------------------------------------------------------------------- 
\subsection{Extractions for objects with spatially varying flux}\label{sec:spatialvar}

\begin{figure*}
\begin{center}
\includegraphics[width=0.99\textwidth]{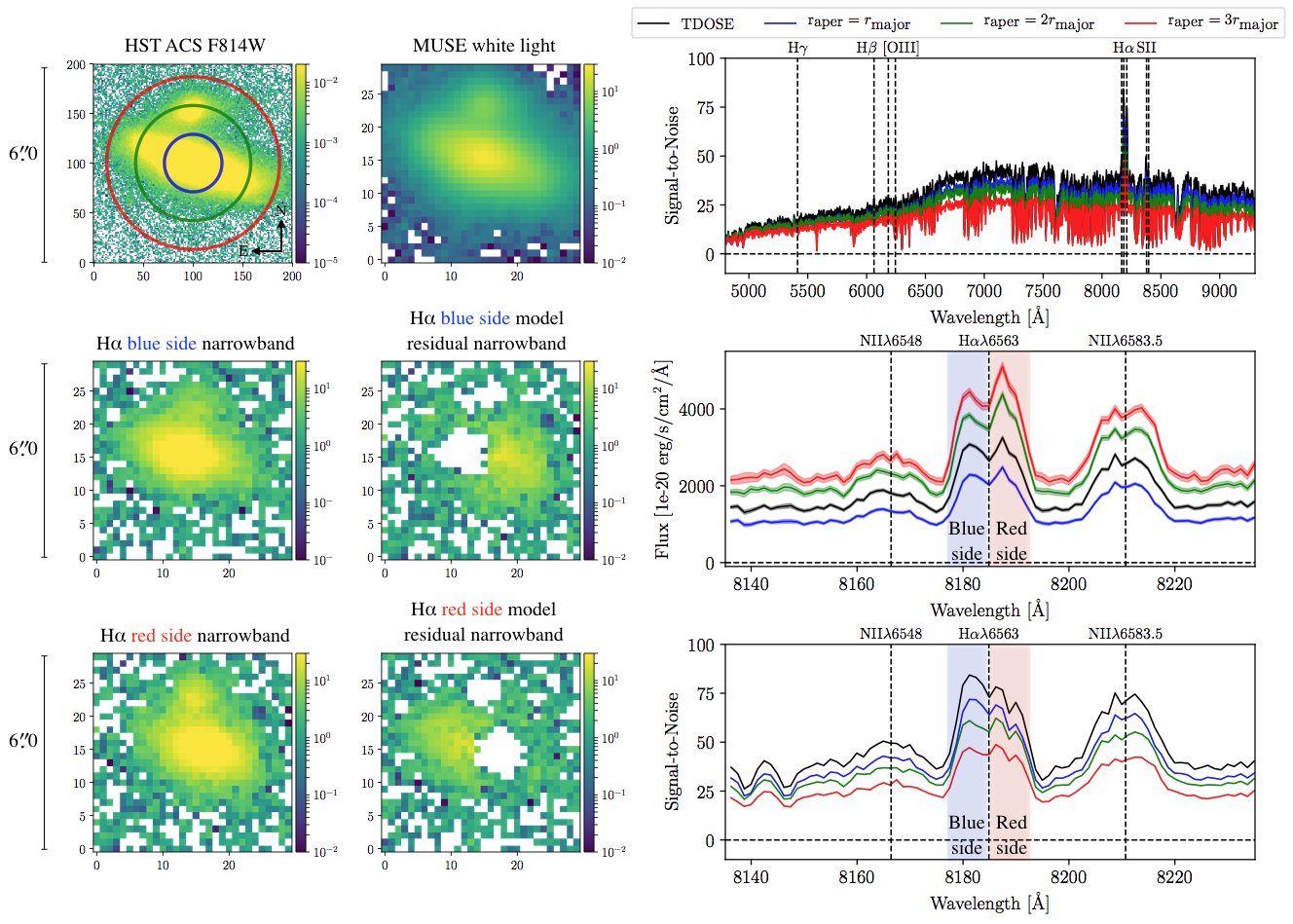}\\
\vspace{-0.4cm}
\caption{Example of spectral extraction of spectra from an object with significant wavelength dependent spatial flux  variation due to rotation.
Such objects provide limitations to the capabilities of extractions with TDOSE based on reference image source models.
The top left panels show the $6\farcs0\times6\farcs0$ postage stamp of the HST ACS F814W image and the MUSE white light image of ID$_\textrm{Guo}=16009$ (ID$_\textrm{MUSE-Wide}=136002114$).
Below these, narrowbands of the blue and red sides of the H$\alpha$ emission in the MUSE data cubes (left) and the residual cube after the removal of the scaled source model (right) are shown.  
The righthand panels show aperture spectra, corresponding to 1 (blue), 2 (green) and 3 (red) $\times r_\textrm{major}$, respectively, together with the TDOSE spectrum based on a single-source Gaussian HST model (black).
The bottom right panels show a narrow zoom-in around the H$\alpha$+NII lines, with the ranges included in the narrowband images in the bottom left panels indicated by the shaded regions.   
It is clear from the over/under-subtraction in the model residual narrowbands, that the single-component TDOSE spectrum is incapable of capturing the shift of the flux in the H$\alpha$ line caused by the galaxy's rotation.
For spectral extractions with prominent spatial shifts, using multi-component reference image models (cf. Figure~\ref{fig:10843}) based on IFS data cube information is recommendable.
}
\label{fig:spatialvar}
\end{center}
\end{figure*}

Even though spectral extractions based on morphological reference image models are generally flexible and versatile in their approach, they do offer some limitations.
As mentioned, spectral extractions with TDOSE assume that the extent of the features to be extracted are well represented by the input model. As explained above, this causes potential biases in representing extended nebular emission if the model reflects the stellar continuum of the galaxy.
Another limitation of the model-based extraction approach occurs when the morphological model represents the source well, but there are spatial variations in the flux distribution as a function of wavelength.
An example of such a spatially varying flux distribution could be strong emission line regions within an otherwise dormant galaxy.
However, such regions are likely identifiable in the reference imaging and can be accounted for based on the reference image model (see, e.g., Figure~\ref{fig:10843}).
A more challenging example of wavelength dependent spatial variations is the presence of velocity shifts of spectral features due to rotation. 
Such effects cannot be identified in (broadband) imaging and can therefore not be corrected for using only reference image modeling.
Figure~\ref{fig:spatialvar} shows the HST F814W image and the MUSE white light image of ID$_\textrm{Guo}=16009$ (ID$_\textrm{MUSE-Wide}=136002114$) in the top left panels.
This object is rotating around an approximate N-S axis which introduces significant shifts of the H$\alpha$ emission.
In the bottom left panels 6\AA{} wide narrow-band filters blue-wards and red-wards of the H$\alpha$ line show the shift of the emission centroid caused by the galaxy's rotation.
In the right panels three aperture spectra (red, green, blue) and a TDOSE spectrum (black) extracted based on a single Gaussian model of the HST reference image are shown for comparison.
The layers included in the blue side and red side H$\alpha$ narrowband images are marked on the spectra in the bottom right panels.
For ID$_\textrm{Guo}=16009$ the TDOSE extraction is biased and is unable to correctly recover the line flux due to the galaxy's rotation.
This is seen by considering the residual narrowbands where the optimized flux model has been subtracted from the blue and red side narrowbands in the bottom central panels.
The single-component HST source model is clearly unable to represent the spatially varying flux distribution of the ID$_\textrm{Guo}=16009$ and causes under- and over-subtraction of the IFS flux around the N-S rotation axis of the galaxy.
To improve the TDOSE model-based extraction of this object, it is not enough to only rely on the information from the reference imaging, where the rotational information is unavailable.
Instead, building a multi-component source model based on the narrowband white light images would provide a much more reliable spectral extraction of ID$_\textrm{Guo}=16009$.
A similar approach is needed to correctly extract spectra with TDOSE of objects with significant wavelength-dependent spatial variations in the overall flux distribution that are not identifiable in the reference imaging.  

% = = = = = = = = = = = = = = = = = = = = = = = = = = = = = = = = = = =
% = = = = = = = = = = = = = = = = = = = = = = = = = = = = = = = = = = =

% ======================================================================
\section{Conclusion}
\label{sec:conc}

The data volume from sensitive wide-field IFSs has grown considerably over the last few decades, 
and with future missions and planned instruments this appears to continue.
Therefore, precise and efficient tools to handle these IFS data cubes, to efficiently de-blend flux in crowded exposures and automatically extract 1D spectra of large samples of objects for further analysis, are needed now, more than ever.
In an attempt to satisfy this demand, we have presented a flexible Python tool for Three Dimensional Optimal Spectral Extraction (TDOSE).

The spectral extraction performed by TDOSE is based on 2D modeling of the object morphology based on (preferably high resolution) reference imaging.
By default, all objects in the FoV are modeled with multivariate Gaussian sources, and are then simultaneously scaled to minimize the difference between the model components and each of the wavelength layers in the IFS data cubes, where the spectra are extracted from.    
This makes TDOSE fully analytical, and hence, very efficient for large samples of objects.
Alternatively, any numerical 2D models, e.g., morphological models from GALFIT or MGE models of individual objects, can be provided to TDOSE as the base for the spectral extraction.

Using MUSE data cubes to illustrate the capability and limitations of TDOSE we show that:
\begin{itemize}
\item The model-based TDOSE extractions are capable of recovering flux from spectra only partially covered by the IFS detector despite a (minimal) loss in S/N caused by the fewer voxels available in the data cube (Section~\ref{sec:specoutside}).

\item Representing objects by multi-source models improves the reliability of the extracted spectra.
For a sample of $\sim$150 [OII] emitters the median increase in flux recovered basing the TDOSE extractions on object models composed of multiple sources as opposed to just one multivariate Gaussian object model, is of the order 5\%. However, in extreme cases, this flux increase can be as high as 50\% for certain wavelength ranges (Section~\ref{sec:multicomp}).

\item The simultaneous scaling of all sources in the FoV allows TDOSE to perform efficient and precise de-blending of sources in 3D.
In this way, TDOSE can be used to remove contaminating (satellite, foreground, or background) sources, when extracting 1D spectra (Section~\ref{sec:deblending}).

\item Spectral extractions performed with TDOSE precisely recover object fluxes comparable to aperture extractions using larger aperture sizes. However, the TDOSE spectra \emph{simultaneously} provide a high S/N, which in the case of aperture extractions is only possible for smaller apertures. Hence, TDOSE brings "the best of two worlds".
The TDOSE extractions for the $\sim$150 [OII] emitters improve the peak [OII] flux and S/N by 9\% and 14\% respectively, compared to aperture extractions (Section~\ref{sec:aperturecomp}).

\item The 3D source models produced by TDOSE are also capable of correcting 2D maps generated from the IFS data cubes, like for instance emission line, metallicity, or kinematic maps (Section~\ref{sec:2Dmaps}).  

\item The main limitation of TDOSE, is that the precision of the spectral extraction is only as good as the model.
For instance, spectral extractions of extended emission based on compact continuum models will naturally be biased. 
In cases where the extent is well-represented by the reference image model, but significant wavelength dependent spatial flux variations are present in the IFS data cube, spectral extractions with TDOSE will also be biased.
However, the flexibility of the multiple source component modeling approach offers several ways to mitigate and account for such biases (Section~\ref{sec:spatialvar}). 
\end{itemize}
Hence, TDOSE offers a flexible, efficient and (by default) fully analytic tool to extract spectra and account for undesirable flux contamination via efficient de-blending in three dimensions from IFS data cubes while optimizing the S/N of the extracted 1D spectra.

% ======================================================================
\begin{acknowledgements}
We would like to thank the MUSE consortium for constructive feedback during the development and testing of TDOSE.
% ---------- GRANTS ----------
This work has been supported by the BMBF grant 05A14BAC % A. Kelz grant 
and we acknowledge support by the Competitive Fund of the Leibniz Association through grant SAW-2015-AIP-2.
% ---------- SOFTWARE ----------
This research made use of the following programs and open-source packages for Python and we are thankful to the  developers of these:
DS9 \citep{2003ASPC..295..489J},
Astropy \citep{2013A&A...558A..33A,2018AJ....156..123A},
APLpy \citep{2012ascl.soft08017R},
iPython \citep{Perez:2007hy},
numpy \citep{vanderWalt:2011dp},
matplotlib \citep{Hunter:2007ux}, 
SciPy \citep{SciPyOpensources:tUcReTVZ}
and
PyFITS which is a product of the Space Telescope Science Institute, which is operated by AURA for NASA.
%

%\NB{Further acknowledgements ?}
\end{acknowledgements}

% WARNING
%-------------------------------------------------------------------
% Please note that we have included the references to the file aa.dem in
% order to compile it, but we ask you to:
%
% - use BibTeX with the regular commands:
%   \bibliographystyle{aa} % style aa.bst
%   \bibliography{Yourfile} % your references Yourfile.bib
%
% - join the .bib files when you upload your source files
%-------------------------------------------------------------------
\bibliographystyle{aa}
\bibliography{paperslibrary}

% ======================================================================
\begin{appendix}
% ======================================================================
\section{Running TDOSE}\label{sec:runningtdose}
% \section{Scripts and Routines}\label{sec:TDOSEscripts}

TDOSE is run through a main setup file that contains the information for the extractions to perform. 
The TDOSE setup file contains pointers to the main inputs for TDOSE, namely the IFS flux data cube, the IFS variance data cube, the reference image of the FoV and a source catalog defining the sources in the FoV.
Among other things, the setup file also defines the extraction mode (aperture, gaussian modeling or pre-defined reference image model), the region of the FoV to consider, what objects to extract spectra for, the PSF model, and the location of initial guesses on the morphological parameters of each source to be modeled (if available). 
A completed setup file is parsed to the main wrapper of TDOSE to perform the spectral extractions.
For extractions from multiple independent IFS data cubes, independent setup files can be generated, and TDOSE can be run in parallel on multiple cores to minimize computation time.
In the following, details are provided on the TDOSE setup file, the main wrappers and routines to use for standard runs of TDOSE and the corresponding outputs produced. 
Appendix~\ref{sec:TDOSErunexamples} presents a few examples of calling sequences and commands for running TDOSE.

Spectral extractions with TDOSE is performed by running \verb+tdose.perform_extraction(setupfile=setupfile)+. 
Here the setup file is the main interface for setting up and controlling the type of spectral extraction TDOSE will perform.
We will describe the setup file in more detail in Section~\ref{sec:setupfile} below.
A range of keywords can be passed to \verb+perform_extraction+ to either skip individual steps or force certain additional features of the extraction. 
For details on these options, we refer the reader to the header of the function itself.
A wrapper around \verb+perform_extraction+ that parallelizes the spectral extractions from multiple data cubes is provided in \verb+tdose.perform_extractions_in_parallel()+.

A successful run of TDOSE will produce a range of outputs which are presented in Section~\ref{sec:TDOSEoutput}. 
The collection of 3D source models are particularly useful for modifications of the data cubes themselves. 
The intrinsic data cubes can be modified with the function \verb+tdose_modify_cube.perform_modification()+ as described below.

In the following, we present a short overview of the main routines and functions of TDOSE, which are called by the \verb+perform_extraction+ function, before we present an overview of the two setup files, and a short description of the outputs that can be generated. 
In Appendix~\ref{sec:TDOSErunexamples} we will provide a few minimal examples for performing standard tasks with TDOSE. 

% - - - - - - - - - - - - - - - - - - - - - - - - - - - - - - - - - - - - - - - - - - - - - - - - - - - - - - - - - - - - 
\subsection{Main TDOSE Scripts}\label{sec:mainscripts}

The file \verb+tdose.py+ contains the main wrapper for performing spectral extractions with TDOSE as it includes the main command to run, namely \verb+tdose.perform_extraction()+. This function calls the main scripts available in the TDOSE repository to carry out the individual task of defining the region to consider, model the reference images, generate the source models, scaling those models to recover the source fluxes at each wavelength range in the data cube, and finally generate the actual 1D spectra as shown in the TDOSE flowchart in Figure~\ref{fig:flowchart}.
The scripts and functions handling these task are described in the following:
\begin{itemize}
\item \verb+tdose.gen_cutouts()+:\\
Before starting the modeling and spectral extraction, this function can be used to generate cutouts of the FoV of relevance from the IFS data cube and the reference imaging in case the full data cube should not be modeled and used for the extraction. 
This limits the required memory needed and makes the modeling and flux scaling more efficient. 
The size of each cutout is defined in the setup file cf. Section~\ref{sec:setupfile}.

\item \verb+tdose.model_refimage()+:\\
This function calls \verb+tdose_model_FoV.gen_fullmodel()+ which performs a multivariate Gaussian modeling of the morphology of all sources in the FoV. Alternatively TDOSE loads a pre-defined reference image model provided by the user.

\item \verb+tdose.define_psf()+: \\
Based on the input from the setup file a 3D wavelength dependent model of the IFS PSF is defined. This PSF model (currently a symmetric 2D gaussian, to keep the PSF convolution analytic), is used to convert the morphological reference image source model to the 3D reference frame of the IFS data.

\item \verb+tdose.model_datacube()+:\\
This function calls \verb+tdose_model_cube.gen_fullmodel()+ which generates a full 3D model of each source based on the reference image model and the model of the IFS PSF. 
This is done by scaling each source component as described in Section~\ref{sec:fluxscaling}.
While scaling each of the individual sources in the model, a collection of the individual de-coupled source model cubes is generated by \verb+tdose_model_cube.gen_source_model_cube()+.
It is this main data structure which is used for de-blending in both 1D and 2D, and for extracting the individual 1D spectra as well as modifying data cubes as described in Sections~\ref{sec:remsources} and \ref{sec:2Dmaps}.

\item \verb+tdose_extract_spectra.py+:\\
This scripts contains the functions used for collapsing the source model cubes to extract the actual 1D spectra. 
The main function \verb+extract_spectra()+ extracts and stores the 1D spectra of individual objects produced by the TDOSE modeling in binary FITS tables. 
A single object spectrum can combine the flux models from multiple sources if desired, and any remaining sources are then treated as contaminants, and accounted for in the spectral extraction.
The extraction is controlled by a "source association dictionary" which defines the sources to assign to each object.
As shown in Appendix~\ref{sec:genspecs} spectra can be generated independent of the main TDOSE extraction by manually generating a source association dictionary and passing it to the function \verb+tdose_extract_spectra.extract_spectra()+.

\item \verb+tdose.plot_spectra()+: \\
If requested in the setup file, the extracted spectra can be plotted in designated wavelength ranges for a quick way to assess the quality of the extractions.
\verb+tdose.plot_spectra()+ is a wrapper calling \verb+tdose_extract_spectra.plot_1Dspecs()+ which provides a flexible plotting and comparison tool which can be used to plot spectra as part of the pist-processing.
E.g., all plotted spectra shown in this paper were generated with \verb+tdose_extract_spectra.plot_1Dspecs()+.
For a list of available keywords and plotting options we refer the reader to the header of the function itself. 
The generated outputs of course allows for further post-processing. 

\item \verb+tdose_utilities.py+:\\
Throughout TDOSE, this script is called, as it continas a collection of useful tools, including, among other things, functions to generate template setup files, building 2D multivariate Gaussians, performing convolutions, preparing modeling with GALFIT, duplicating setup templates, generate overview plots and extracting sub-regions of images data cubes.

\item \verb+tdose_build_mock_cube.py+:\\
For testing and trouble shooting a small package of functions to generate mock data cubes is provided in this script.

\item \verb+tdose_modify_cube.py+:\\
This is the last main script available in the TDOSE repository. 
As mentioned above, this is used to modify the input 3D data cube, by subtracting the 3D source models generated by TDOSE to illiminate undesired flux contamination. 
This modification is controlled by the setup file \verb+tdose_setup_template_modify.txt+ described in Appendix~\ref{sec:setupmodify} below.
A template of this file can be generated with the function \verb+tdose_utilities.generate_setup_template_modify()+ and is provided with the TDOSE version 3.0 release.

\end{itemize}

% - - - - - - - - - - - - - - - - - - - - - - - - - - - - - - - - - - - - - - - - - - - - - - - - - - - - - - - - - - - - 
\subsection{The TDOSE Setup File}\label{sec:setupfile}

The TDOSE setup file is the main way to interact and set up a spectral extraction with TDOSE and is an input for \verb+tdose.perform_extraction()+.
A template setup file (\verb+tdose_setup_template.txt+) is provided in the TDOSE Github repository at \url{https://github.com/kasperschmidt/TDOSE} (and the packed TDOSE version 3.0 release) and can be generated with the \verb+tdose_utilities.generate_setup_template()+ function.
The setup file contains several sections with different information on both inputs and outputs. 
In the following, each of these are described. 
For further details, the template setup file itself provide comments on each of the input values.

\begin{itemize}
\item \verb+DATA INPUT+:\\
The data input section defines the input required for a minimal default TDOSE run.
Here the location of the IFS data cube to extract the spectra from and the corresponding variance cube, used for estimating the uncertainty on the extracted spectra, are provided.
This section also points to the reference image as well as the location and column names of the main source catalog.
The source catalog defines which sources to model (assuming TDOSE is responsible for generating the reference image source model). It can be generated with standard source detection softwares, or be put together manually if more detailed modeling should be performed.
A weight image of the reference image can also be provided. This is not used by TDOSE itself, but if only a sub-region of the IFS data cube is considered, it can be useful to have the relevant FoV cut out of the reference weight image together with the reference image.
Hence, any image can actually be provided here.
For instance is can be useful to have TDOSE cut out from a "sigma image" if GALFIT modeling of the reference image cutout will be performed.

\item \verb+OUTPUT DIRECTORIES+:\\
This section defines the location of the outputs generated with TDOSE.

\item \verb+CUTOUT SETUP+:\\
Here the cutouts around each source in the source catalog to consider during the extraction and modeling are defined. It is recommended to always use a cut out for the extractions as this both limits the required memory available, and speeds up calculations. 
The cutout sizes can be defined to have the same size for all objects by providing the dimensions in R.A. and Dec. given in arc seconds. Alternatively, the location of a file with source specific cut out sizes can be provide here. 

\item \verb+SOURCE MODEL SETUP+:\\
The source model setup defines what extraction method to use.
The currently enabled modes which are also shown in Figure~\ref{fig:flowchart} are:

\subitem \verb+gauss+: 
This is the default mode of TDOSE. Here, the reference image will be modeled by positioning a multivariate Gaussian component at the location of each source in the (cutout) FoV. 
Using this reference image model makes the spectral extraction fully analytic.
The optimization of these morphological models is performed using Scipy's \verb+curve_fit+ \citep[\url{https://www.scipy.org};][]{SciPyOpensources:tUcReTVZ} function.    
The details of this extraction is provided in the \verb+GAUSS MODEL SETUP+ section of the setup file described below.

\subitem \verb+modelimg+: 
This option allows the user to provide an already existing model of the reference image, as detailed in the \verb+MODEL IMAGE SETUP+ section of the setup file, instead of using the build-in Gaussian modeling of TDOSE. 
Such models can for instance be generated with GALFIT.
This model will be treated numerically as opposed to the Gaussian model generated by TDOSE.
If only a model image is provided, TDOSE assumes that it represents the 1 object of interest in the FoV.
To be able to de-blend sources when extracting spectra a model cube needs to be provided as specified under the \verb+MODEL IMAGE SETUP+ section (see below).

\subitem \verb+aperture+:
This option performs aperture extractions as described in Section~\ref{sec:PSFandAperureExtractions}. 
The aperture sizes for this mode are provided in the  \verb+APERTURE MODEL SETUP+ section of the setup file described below.

\item \verb+GAUSS MODEL SETUP+:\\
If the default \verb+gauss+ extraction is performed, it is possible to provide initial estimates of the relative sizes and orientation of each source in the source catalog, in this section of the setup file.
This is done by providing a SExtractor \citep{1996A&AS..117..393B} output with gaussian morphological parameter estimates and flux scalings. This enables a quicker convergence of the reference image modeling of TDOSE.
This section also provides a limit on how much the source catalog source position and the centroid of the corresponding multivariate Gaussian model are allowed to differ.
This is particularly important when trying to model faint objects, as in these cases TDOSE might attempt to fit noise, and the model location might be fixed on noise peaks at random location in the FoV as opposed to the peak flux close to the location of the source.

\item \verb+MODEL IMAGE SETUP+:\\
If a \verb+modelimg+ extraction was chosen this part of the setup file provides the path to the directory containing the individual source models.
TDOSE will look for a model named as the reference image (cut out) it is suppose to represent with "\verb+model_+" prepended. If no model is found with this name the object is skipped.
If a model appended \verb+_cube+ is found in the directory, it is assumed that this file contains a cube with the individual model components isolated in individual layers of the cube. 
A model cube will always have priority over a model image.
Based on the parent IDs in the source catalog, the model cube will be used to define what sources belong to the object of interest (i.e., to extract a spectrum for), and what sources should be considered contaminants. 
GALFIT models can be converted to TDOSE-suited model-cubes with \verb+tdose_utilities.galfit_convertmodel2cube()+.
If individual isolated source model images are available, these can be assembled into a  TDOSE-suited model-cube with \verb+tdose_utilities.build_modelcube_from_modelimages()+.
For examples see Appendix~\ref{sec:modelimgext}.

\item \verb+APERTURE MODEL SETUP+:\\
Here the sizes of the apertures to use for an aperture extraction are provided. 
A single number, a list, or the location of a text file with source specific apertures can be provided. 

\item \verb+PSF MODEL SETUP+:\\
This section of the setup file defines the IFS PSF and its wavelength dependence.
TDOSE version 3.0 focuses on analytic spectral extraction and therefore currently only allows a symmetric 2D Gaussian representation of the IFS PSF.
The wavelength dependence of the PSF is described as a linear evolution on the form  
\begin{equation}
\textrm{FWHM}(\lambda) = p_0[''] + p_1[''/\textrm{\AA}] \times (\lambda - p_2[\textrm{\AA}])
\end{equation}
where each of the parameters $p_0$, $p_1$ and $p_2$ have to be provided.
This follows the Gaussian PSF description of the MUSE-Wide fields described in Table~2 of \cite{2019A&A...624A.141U} where $p_2=7050$\AA.

\item \verb+NON DETECTIONS+:\\
If any of the sources listed in the source catalog, e.g., faint sources that are hard to model reliably, or sources with emission lines detected in the IFS data without clear continuum counterparts, should be treated as point sources, they can be specified in this section. 
Either a list of IDs or a text file listing the sources to treat as non-detections can be provided.
If the source model is set to the default \verb+gauss+ option, the sources listed will be replaced in the reference image models by a single point source (within the radius of ignorance which is also provided here) before convolution with the IFS PSF.
If on the other hand the source model mode is \verb+modelimg+, TDOSE assumes that the provided reference image model (cube) already represents the desired extraction of the non-detection, and therefore ignores the info provided in the \verb+NON DETECTIONS+ section of the setup file

\item \verb+CUBE MODEL SETUP+:\\
If the user is only interested in a specific part of the wavelength range of the IFS data cube, the layers to model and output can be specified here. By default, TDOSE models and extracts spectra from all wavelength layers.
This part of the TDOSE setup file also provides the name scheme for the outputs to generate and defines the source model optimizer to use. The one described in Section~\ref{sec:fluxscaling} which is the standard of TDOSE is selected as \verb+matrix+. 
Optimization can also be done using \verb+Scipy+'s non-negative least squares solver by selecting \verb+nnls+.
In this case the flux scales $\mathbf{a}_m$ described in Section~\ref{sec:fluxscaling} are restricted to be $\ge0$. 
As mentioned $\mathbf{a}_m$ describes the spectrum of a given source, hence, for bright sources with continuum (and modest absorption), formally $\mathbf{a}_m\ge0$ should always be true. 
Due to noise in the IFS data, this is however not true for fainter sources, where scales can be (slightly) negative, in which case the more general \verb+matrix+ optimization is preferred.

\item \verb+SPECTRAL EXTRACTION+:\\
If only a sub-set of the sources in the source catalog are of scientific interest (the remaining being considered as contaminating sources), there is no reason to extract spectra of all the sources in the source catalog.
Through a list of source catalog IDs (or a text file with IDs) the sources to extract spectra for are defined.
This section also defines the prefix used in naming the output spectra.

\item \verb+PLOTTING+:\\
The last part of the TDOSE setup file describes the wavelength, flux and S/N ranges to plot for the extracted spectra.  
If plotting is requested an overview plot for each object is also generated.
\end{itemize}

% - - - - - - - - - - - - - - - - - - - - - - - - - - - - - - - - - - - - - - - - - - - - - - - - - - - - - - - - - - - - 
\subsection{The TDOSE Modification Setup File}\label{sec:setupmodify}

A key feature of the TDOSE source model cubes that can be outputted when extracting spectra, is that they can be used to subtract undesired sources and the corresponding fluxes in full 3D from the input IFS data cube.
This is beneficial for removing and/or de-blending flux of different sources, when generating 2D images and maps from the data cube as described in Sections~\ref{sec:remsources} and \ref{sec:2Dmaps}.
The TDOSE modification setup file provides a simple way to perform such an IFS data cube modification.
All that is required to complete the modification setup file is the location and name of the data cube, the collection of 3D source models generated by TDOSE, a pointer to the output directory of the modified cube, and a list of the sources to remove in the original data cube.
The setup file is then passed to \verb+tdose_modify_cube.perform_modification()+ to correct the data cube.

% - - - - - - - - - - - - - - - - - - - - - - - - - - - - - - - - - - - - - - - - - - - - - - - - - - - - - - - - - - - -  
\subsection{TDOSE Output}\label{sec:TDOSEoutput}
This section describes the main outputs of a TDOSE spectral extraction. 
Several of them are referred to in the description of the individual scripts and functions above.
In "[]" after each output listed below, we provide the parameter in the TDOSE setup file (Section~\ref{sec:setupfile}) that defines the name extension of the output to ease the identification of the outputs.
The main TDOSE outputs are:
\begin{itemize}
\item Reference Image Model [\verb+model_image_ext+]:\\
A 2D FITS Image containing a model of the reference image (cutout) if TDOSE is asked to model the sources. 
It is this model that defines the morphology of the sources in the source catalog, before convolution with the PSF is performed. 
If the source model mode is \verb+gauss+ this model is generated by fitting multivariate Gaussians to each of the sources described in the source catalog.

\item Source Model Parameters [\verb+model_param_reg+]:\\
A DS9 \citep{2003ASPC..295..489J} region file marking the location and extent of the Gaussian model parameters describing the individual source models used to generate the reference image model.
This can be useful for assessing the quality of the reference image model and for locating each individual model in the FoV.

\item Image Model in the IFS Frame [\verb+model_image_cube_ext+]:\\
A 2D Fits Image showing the reference image model in the IFS data cube "reference frame", i.e., after converting the pixel scales of the reference image to match the IFS data cube.

\item PSF Cube [\verb+psf_savecube+]:\\
A 3D FITS data cube, containing the IFS PSF model defined in the TDOSE setup file. 
The PSF is spatially centered and captures the wavelength evolution. 
If this cube is saved, it will be named by prepending \verb+_psfcube_+ to the FITS file containing the source model cubes.

\item Data Cube Model [\verb+model_cube_ext+]:\\
This output is again a 3D FITS data cube. 
It contains the model data cube of the IFS data cube after convolution of the reference image model with the IFS instrument PSF and after each of the individual source components in the model have been flux-optimized to match the IFS data cube flux levels at each wavelength layer in the data cube as described in Section~\ref{sec:fluxscaling}.

\item Residual Data Cube Model [\verb+residual_cube_ext+]:\\
A 3D FITS data cube providing the residual, i.e., the difference between the original IFS data cube and the 3D data cube model.
The cube contains "IFS data cube" - "data cube model".

\item Source Model Cubes [\verb+source_model_cube_ext+]:\\
A FITS data structure containing the de-blended 3D source models for all $n$ modeled sources in the FoV.
This can be used to account for individual sources in full 3D and modify the input IFS data cube accordingly by removing individual source models with \verb+tdose_modify_cube.py+.
The dimensions of the FITS structure is [source number, wavelength, y-axis, x-axis]. 
Hence, the last three dimensions define the 3D model for the chosen source number.

\item Extracted 1D Spectra [\verb+spec1D_name+]:\\
The extracted spectra are generated by collapsing the source model cubes of the sources contributing to the object of interest as described in Section~\ref{sec:extraction}.
The indvidual spectra are provided as multi-extension FITS files containing a binary table with the fluxes, flux errors, S/N and wavelength of the extracted spectrum, and the 3D object model used for the extraction. 
As described in Section~\ref{sec:extraction}, an object spectrum can be an arbitrary combination of any $k$ sources in the FoV. However, by default each object corresponds to a single source in the source catalog, and the number of sources therefore equals the number of extracted object spectra if all spectra are extracted.

\item Plots of the 1D Spectra [\verb+plot_generate+]:\\
Lastly, if requested TDOSE generates plots of the flux and S/N spectra extracted and an overview plot for each object extracted.

\end{itemize}
The command line output of TDOSE assembles copy-paste ready DS9 \citep{2003ASPC..295..489J} commands, that are useful for displaying and inspecting the various FITS data cubes and images generated and used in a TDOSE extraction.
To take advantage of these commands a functioning command line version of DS9 must be available.

% ======================================================================
\section{Running TDOSE - a few examples}\label{sec:TDOSErunexamples}
In the following subsections a few examples of calling sequences and commands to execute the spectral extractions and other tools provided as part of the TDOSE software package are presented.
The shown examples of Python code can be copy-pasted into for instance iPython after updating the inputs required.
For keywords and parameters available beyond the ones described here, we refer to the docstring of the individual functions and procedures which (accessible by for instance typing \verb+? tdose.perform_extraction()+ in iPython).

% - - - - - - - - - - - - - - - - - - - - - - - - - - - - - - - - - - - - - - - - - - - - - - - - - - - - - - - - - - - - 
\subsection{TDOSE $\mathtt{gauss}$ Extractions}\label{sec:gaussext}
This extraction mode is the default spectral extraction with TDOSE and can be performed with 
\begin{lstlisting}
import tdose
# --- INPUT ---
setupfile = '/Path/to/setupfile/tdose_setup_template.txt'
# --- COMMAND ---
tdose.perform_extraction(setupfile=setupfile,performcutout=True,generatesourcecat=True,verbose=True,verbosefull=True)
\end{lstlisting}
Here the \verb+source_model+ parmeter in the setup file should be set to \verb+gauss+

TDOSE will position multivariate Gaussians at the position of all sources in the source catalog when modeling the reference image. 
Such a source catalog can be generated with standard methods like SExtractor, or be generated by hand.
In the latter case, a simple ascii file can be converted to a fits catalog with
\begin{lstlisting}
import tdose_utilities as tu
# --- INPUT ---
outputdir = '/Path/to/output/directory/'
catfile = outpath+'manually_generated_source_catalog.txt'
# --- COMMAND ---
outputfile = tu.ascii2fits(catfile,asciinames=True,skip_header=7,outpath=outputdir,verbose=True)
\end{lstlisting}

It can be useful to point to a SExtractor photometric catalog with morphological estimates with the parameter \verb+gauss_guess+ in the setup file. 
These estimates will then be used as initial guesses for the Gaussian modeling of the FoV and will likely improve the precision of the resulting model. 
In case sources were added by hand to the source catalog, a default gaussian point source will be used unless these have shape measurements added to the SExtractor catalog.

% - - - - - - - - - - - - - - - - - - - - - - - - - - - - - - - - - - - - - - - - - - - - - - - - - - - - - - - - - - - - 
\subsection{TDOSE $\mathtt{modelimg}$ Extractions}\label{sec:modelimgext}

To base the TDOSE extraction on an existing model of the reference image instead of having TDOSE generate a model of multivariate Gaussians the extraction mode \verb+modelimg+ can be used.
Simply updating the keywords \verb+sourcecatalog+ (if different for provided model), \verb+source_model+ and \verb+modelimg_directory+ and re-running the \verb+tdose.perform_extraction()+ command as described above, will extract spectra based on the models found in the model directory.
TDOSE expect to either find models of the reference image (named as the reference image prepended \verb+model_+) or a cube containing individual source models (named as the reference image model appended \verb+_cube+).
The latter model format is necessary for performing de-blending for \verb+modelimg+ extractions.

\subsubsection{Using GALFIT Models}
The \verb+tdose_utilities.py+ script includes a selection of tools to handle GALFIT models consisting of S\`ersic and Gaussian components. 
Most importantly, to perform de-blending using a GALFIT model a cube of the individual GALFIT components which is compatible with TDOSE can be generated with:
\begin{lstlisting}
import tdose_utilities as tu
# --- INPUT ---
import astropy.io.fits as afits
galfitmodel = '/Parth/to/GALFIT/model/model_ref_image.fits'
sourcecat_compinfo = '/Path/to/source/component/info/model_componentinfo.txt'
PSFkernel = afits.open('/Path/to/reference/image/PSF/model/PSFmodel')[0].data
# --- COMMAND ---
tu.galfit_convertmodel2cube([galfitmodel],savecubesumimg=True,includewcs=True,convkernels=[PSFkernel],sourcecat_compinfo=sourcecat_compinfo)
\end{lstlisting}
Here the optional PSF kernel input is used to convolved the reference image model if not already accounted for. 
The component info file associates individual sources to objects in the model. TDOSE expects an ascii file containing the file model name, the object ID and a designation of which sources in the 3D cube that should belong to the object and which should be counted as contaminants. This is indicated by strings on the format 'X:Y' where X counts the sources (starting from 1 corresponding to the \verb+COMP_X+ GALFIT model header keyword) and Y indicates whether a model component belongs to the object (Y=1), is a contaminant (Y=2) or represents the sky model (Y=3).
Hence, the following indicates a model with model components 1 and 3 belonging to the main object (ID=55), whereas sources 2 and 4 will be treated as contaminants when TDOSE de-blends the model components. The 5th model component represents the sky:
\begin{verbatim}
model_ref_image.fits  55  1:1  2:2  3:1  4:2  5:3
\end{verbatim}     

\subsubsection{Using Independent (MGE) Source Models}
Alternative to using GALFIT models, individual 2D source models of individual sources can be combined to a cube of source models with the commands
\begin{lstlisting}
import tdose_utilities as tu, glob, numpy as np
# --- INPUT ---
models2D    = glob.glob('/Path/to/2D/models/*.fits')
modelsext   = np.ones(len(models2D)) # list of FITS extensions with models
basename    = '/Path/to/output/and/base/naming/combined_models'
# --- COMMAND ---
tu.build_modelcube_from_modelimages(models2D,modelsext,basename,savecubesumimg=True)
\end{lstlisting}
To perform de-blending for this approach, the individual de-blended spectra can be extracted based on the 3D source models generated by TDOSE using a source association dictionary as described in Appendix~\ref{sec:mainscripts} and \ref{sec:genspecs}.

% - - - - - - - - - - - - - - - - - - - - - - - - - - - - - - - - - - - - - - - - - - - - - - - - - - - - - - - - - - - - 
\subsection{TDOSE $\mathtt{aperture}$ Extractions}

TDOSE is also capable of extracting aperture spectra. 
To do this the setup file parameter \verb+source_model+ should be set to \verb+aperture+ and the aperture size(s) to use for each object should be provided via the \verb+aperture_size+ parameter. 
With these parameters set the extraction can be performed by running the commands provided in Appendix~\ref{sec:gaussext} above.

% - - - - - - - - - - - - - - - - - - - - - - - - - - - - - - - - - - - - - - - - - - - - - - - - - - - - - - - - - - - - 
\subsection{Generate large number of setup files for multiple extractions}

Extractions from a large number of data cubes is handled by running multiple instances of the \verb+tdose.perform_extraction()+ function (potentially in parallel with \verb+tdose.perform_extractions_in_parallel()+).
Individual setup files will handle the individual extractions.
Handling and editing a large number of TDOSE setup files can be done with:  
\begin{lstlisting}
import tdose_utilities as tu
# --- INPUT ---
outputdir = '/Path/to/output/directory/'
infofile = outputdir+'tdose_setupfile_info.txt'
namebase = 'tdose_setupfile_namebase'
# --- COMMAND ---
tu.duplicate_setup_template(outputdir,infofile,namebase=namebase,loopcols='all')
\end{lstlisting}
Here the infofile is a simple ascii file containing the values of the parameters to edit in the individual setup files. The first column named \verb+setupname+ indicates what to append to the \verb+namebase+ when naming the setup files.
The rest of the columns provide the information for each of the setup file parameters (columns should be named according to these) to replace in the template setup file (generated with \verb+tdose_utilities.generate_setup_template()+).
Hence, all setup files can be generated and edited with just a single file.

% - - - - - - - - - - - - - - - - - - - - - - - - - - - - - - - - - - - - - - - - - - - - - - - - - - - - - - - - - - - - 
\subsection{Generate 1D Spectra from 3D Source Models}\label{sec:genspecs}

The collection of 3D source models produced and outputted by TDOSE form the basis of the spectral extraction of individual objects.
As described in Section~\ref{sec:extraction} objects are extracted by combining and collapsing the scaled models of one or more sources from the source catalog.
If multiple sources are extracted from the same data cube, TDOSE centers the cut out region on each source and perform the extraction.
However, for some applications it can be useful to extract multiple objects from a single model without re-centering the FoV. 
This can be done by running the extraction tool on an existing FITS structure containing the 3D source models cubes: 
\begin{lstlisting}
import tdose_extract_spectra as tes
# --- INPUT ---
data_cube_file          = '/Path/to/data/datacube.fits'
data_cube_ext         = 'DATA'
variance_cube_file   = '/Path/to/data/variancecube.fits'
variance_cube_ext   = 'VAR'
model_cube_file       = '/Path/to/data/datacube_tdose_modelcube_gauss.fits'
model_cube_ext       = data_cube_ext
sourcemodels_file                      = '/Path/to/data/datacube_tdose_source_modelcube_gauss.fits'
sourcemodels_ext                      = data_cube_ext
nameextension  = 'tdose_spectrum_manual_extract'
outputdir          = '/Path/to/output/directory/'
SAD                 = {111:[0,4,5], 222:[2], 333:[1,6,7]}
# --- COMMAND ---
specfiles  = tes.extract_spectra(model_cube_file,model_cube_ext=model_cube_ext,nameext=nameextension, source_association_dictionary=SAD, outputdir=outputdir, variance_cube_file=variance_cube_file, variance_cube_ext=variance_cube_ext,source_model_cube_file=sourcemodels_file,source_cube_ext=sourcemodels_ext, data_cube_file=data_cube_file, verbose=True)
\end{lstlisting}
Here \verb+SAD+ is the source association dictionary associating sources to individual objects as described in Appendix~\ref{sec:mainscripts} above. In the above example object 111 is comprised of source 0, 4 and 5, object 222 corresponds to source 2 and object 333 is comprised of source 1, 6 and 7 from the reference image model.

% - - - - - - - - - - - - - - - - - - - - - - - - - - - - - - - - - - - - - - - - - - - - - - - - - - - - - - - - - - - - 
\subsection{Modify Data Cubes Using 3D Source Models}\label{sec:modcubes}

The collection of 3D source models produced by TDOSE enable correcting the intrinsic flux data cubes for contamination in full 3D as described in Section~\ref{sec:remsources} and shown in Section~\ref{sec:2Dmaps}.
Modification of original IFS data cubes and thereby removing contaminating flux based on the TDOSE source models is controled by the modification setup file described in Appendix~\ref{sec:setupmodify} above. 
The modification is performed with
\begin{lstlisting}
import tdose_modify_cube as tmc
# --- INPUT ---
setupfile = '/Path/to/modify/setupfile/tdose_setup_template_modify.txt'
# --- COMMAND ---
tmc.perform_modification(setupfile=setupfile)
\end{lstlisting}

% ======================================================================
\end{appendix}
% ======================================================================
\end{document}